\documentclass[conference]{IEEEtran}
\usepackage[utf8]{inputenc}
\usepackage[english]{babel}
\usepackage{graphicx}
\usepackage[caption=false]{subfig}
\usepackage{stmaryrd}
\usepackage{amsmath}
\usepackage{amssymb}
\usepackage{amsfonts}
\usepackage{amsthm}
\usepackage{listings}
\usepackage{color}
\usepackage{proof}
\usepackage{mathpartir}
\usepackage{thm-restate}
\usepackage{comment}
\usepackage{mathtools}
\usepackage{framed}
\usepackage{multicol}
\usepackage{hyperref}
\usepackage{dashbox}
\usepackage{url}
\usepackage{placeins}
\usepackage{enumitem}
\usepackage{xcolor}
\usepackage[numbers,sort]{natbib}

\theoremstyle{plain}

\newtheorem{lemma}{Lemma}

\theoremstyle{definition}
\newtheorem{definition}{Definition}

\newcommand{\case}[1]{\textbf{Case} #1\textbf{:}}
\newcommand{\basecase}[1]{\textbf{Base case} #1\textbf{:}}
\newcommand{\inductivecase}[1]{\textbf{Inductive case} #1\textbf{:}}

\newcommand{\inenv}{I}
\newcommand{\outenv}{O}

\newcommand{\packetqueue}{\vec{q}}
\newcommand{\schedule}{\pi}
\newcommand{\timestamp}{\textit{ts}}

\newcommand{\inttype}{\textbf{int}}
\newcommand{\stringtype}[1]{\textbf{string}_{#1}}

\newcommand{\pc}{\textit{pc}}

\newcommand{\ltype}[2]{#1 @ #2}

\newcommand{\ladv}{\ell_\textit{adv}}
\newcommand{\ladvequiv}{\approx_{\ladv}}
\newcommand{\ladvequivnet}{\approx^{\textit{net}}_{\ladv}}
\newcommand{\ladvequivprog}{\approx^{\sbullet[1.2]}_{\ladv}}

\newcommand{\assignevent}[2]{\textsf{a}(#1,#2)}
\newcommand{\assignsizeevent}[2]{\textsf{|a|}(#1,#2)}
\newcommand{\queueevent}[2]{\textsf{q}(#1,#2)}
\newcommand{\scheduleevent}[3]{\textsf{s}(#1,#2,#3)}
\newcommand{\inputevent}[3]{\textsf{i}(#1,#2,#3)}
\newcommand{\outputevent}[2]{\textsf{o}(#1,#2)}

\newcommand{\xrect}[1]{
    \setlength{\fboxsep}{1.3pt}\setlength\fboxrule{0.75pt}
    \boxed{\ensuremath{\phantom{v}\mkern-9mu #1}}
}

\newcommand{\xangled}[1]{\langle #1 \rangle}
\newcommand{\xAngled}[1]{\llangle #1 \rrangle}

\newcommand{\lconfig}{P}
\newcommand{\eval}[3]{\xangled{#1,#2} \Downarrow #3}

\newcommand{\rightarrowdbl}{\rightarrow\mathrel{\mkern-14mu}\rightarrow}

\newcommand\sbullet[1][.5]{\mathbin{\vcenter{\hbox{%
  \scalebox{#1}{$\bullet$}}}}%
}

\newcommand{\projevent}[2]{\lfloor #1 \rfloor_{#2}}
\newcommand{\projeventinternal}[2]{\lfloor #1 \rfloor^{\sbullet[1.2]}_{#2}}
\newcommand{\filtertrace}[2]{#1 \upharpoonright #2}
\newcommand{\filtertraceinternal}[2]{#1 \upharpoonright^{\sbullet[1.2]} #2}

\definecolor{backgroundColor}{rgb}{1.0, 0.95, 0.95}
\definecolor{commentsColor}{rgb}{0.35, 0.35, 0.35}
\definecolor{keywordsColor}{rgb}{0.0, 0.0, 0.65}
\definecolor{stringColor}{rgb}{0.5, 0.0, 0.15}

\usepackage{lmodern}

\usepackage{stix}

\def\istechnicalreport{} 

\title{\huge Towards Language-Based Mitigation of Traffic Analysis Attacks}
\author{
    \IEEEauthorblockN{Jeppe Fredsgaard Blaabjerg}
    \IEEEauthorblockA{Aarhus University\\jfblaa@cs.au.dk}
    \and
    \IEEEauthorblockN{Aslan Askarov}
    \IEEEauthorblockA{Aarhus University\\aslan@cs.au.dk}
}

\begin{document}
\thispagestyle{empty}
\setcounter{page}{1}
\pagestyle{plain}

\maketitle






\begin{abstract}
    Traffic analysis attacks pose a major risk for online security. Distinctive patterns in communication act as fingerprints, enabling adversaries to de-anonymise communicating parties or to infer sensitive information.
    Despite the attacks being known for decades, practical solution are scarce. Network layer countermeasures have relied on black box padding schemes that require significant overheads in latency and bandwidth to mitigate the attacks, without fundamentally preventing them, and the problem has received little attention in the language-based information flow literature. Language-based methods provide a strong foundation for fundamentally addressing security issues, but previous work has overwhelmingly assumed that interactive programs communicate over secure channels, where messages are undetectable by unprivileged adversaries. This assumption is too strong for online communication where packets can be trivially observed by eavesdropping.
    In this paper we introduce SELENE, a small language for principled, provably secure communication over channels where packets are publicly observable, and we demonstrate how our program level defence can reduce the latency and bandwidth overheads induced compared with program-agnostic defence mechanisms. We believe that our results constitute a step towards practical, secure online communication.
\end{abstract}


\section{Introduction} \label{section:introduction}


Work on traffic analysis attacks has shown that many systems and services are vulnerable to de-anonymisation and loss of secrecy by producing distinctive patterns in their network traffic.
Traffic analysis has particularly been studied in the context of anonymous communication and website fingerprinting \cite{das-meiser-mohammadi-kate-2017,cherubin-2017,juarez-2015,kwon-2015,overdorf-2017,panchenko-2015,siby-2018}.
Defence strategies against website fingerprinting are commonly done at the network level \cite{cherubin-2017}, and rely on constant rate padding, where source traffic is morphed to fit a predefined target pattern \cite{fu-2003}. 
Constant rate padding can be applied in a black box fashion, making it an intuitively appealing technique against website fingerprinting. However, it often falls short in practice as achieving a high degree of security introduces intolerable bandwidth and latency overheads for many applications \cite{dyer-2012}, such as anonymous, low-latency browsing and communication \cite{feigenbaum-2010} and privacy preserving IoT devices \cite{apthorpe-2018}. Cherubin et al. argue that the application layer defences against traffic analysis are more natural as they act directly on the objects that are fingerprinted at the network level, while defences at lower layers must model legitimate traffic in order to generate convincing traffic padding \cite{cherubin-2017}.

Website fingerprinting is not the only attack made possible by traffic analysis.
Online services that process sensitive information are now ubiquitous and previous work has shown that many such services are vulnerable to attack, as their communication behaviour reveals system secrets.
Analysis by Chen et al. suggests that the scope of the issue is industry-wide \cite{chen-2010}. Their study finds that design features used for creating reactive sites generate characteristic traffic patterns that allow an adversary to infer highly detailed, sensitive user information. They demonstrate this vulnerability across a number of high-profile websites, e.g. they are able to infer which illness a user selects on an online health site, and argue that traffic analysis attacks pose an unprecedented threat to the confidentiality of user information processed by online systems, and that this information is often far more sensitive than identifying which website a user visits as studied in anonymity research.

Language-based information flow methods provide principled ways of enforcing that the observable behaviour of a program does not depend on secrets. The language-based approach is appealing as the security condition of noninterference \cite{goguen-meseguer-1982} can be provably enforced using a type system. O'Neill et al. formulate a noninterference condition for interactive programs \cite{oneill-clarkson-chong-2006}, where programs communicate over in- and output channels. Their condition requires that input on secret channels does not influence output on public channels. This condition has been used in a breadth of other work \cite{schoepe-sabelfeld-2015,bastys-2020,bohannon-2009,askarov-sabelfeld-2009,clarkson-scheider-2010,clark-hunt-2009,hedin-sabelfeld-2012}. Unfortunately, the models used in these works assume that messages on secret channels are invisible to adversaries. Other work models Internet communication using expressly public channels \cite{mantel-sabelfeld-2003}.
This makes the security results inapplicable for reasoning about online services and distributed programs where secret information is shared between remote, trusted entities.
The only other work we are aware of, that allows an adversary to observe the communication behaviour of a program on non-public channels, is by Sabelfeld and Mantel \cite{sabelfeld-mantel-2002}. They consider encrypted channels that protect message contents, but do not hide message presence, and give a timing-sensitive security condition. Their security condition does not consider the size of messages, which can be exploited in traffic analysis attacks, and their semantics lets the blocking behaviour of receives on encrypted channels be public by letting number of available messages be public.

Even simple interactions are not secure when messages can be eavesdropped. We demonstrate this using four example programs that each highlight a different source of leaking. In the following examples we consider a simple two point lattice with elements $\{\texttt{L},{\texttt{H}}\}$ and ordering $\texttt{L} \sqsubseteq \texttt{L}$, $\texttt{L} \sqsubseteq \texttt{H}$, and $\texttt{H} \sqsubseteq \texttt{H}$, and adopt the convention that low variables start with $\texttt{l}$ and high variables start with $\texttt{h}$. For simplicity, we assume that $\texttt{Public}$ is a channel at level $\texttt{L}$ and all other channels, e.g. $\texttt{Alice},\texttt{Bob}$, are at level $\texttt{H}$.
We first consider a program, where a number of messages are sent depending on the value of a secret variable:
\begin{lstlisting}[captionpos=none]
    /* Program 1 - Message count */
h_count = 0;
while (h_count < h_secret)
do {
    out(Alice,1);
    h_count = h_count + 1;
}
\end{lstlisting}
Program 1 satisfies the common security conditions of previous works, as the value of secret variable \texttt{h\_secret} only influences secret output (line 5). However, if traffic can be eavesdropped, the program is trivially insecure. By counting the number of messages sent, an adversary can easily infer the exact value of the secret, as the number of messages depends on the secret.

We assume that each message is tagged with recipient information and that the adversary can observe both the size of each message and the time at which it was sent, i.e., the adversary is timing-sensitive.
These assumptions lead us to naturally identify three other sources of leaks exemplified by the following programs, where respectively the recipient, the size, and the timing of messages leak secrets. These programs would commonly be considered secure in previous work.
\begin{lstlisting}[captionpos=none]
    /* Program 2 - Recipient of message */
if (h) then {
    out(Alice, 42);
} else {
    out(Bob, 42);
}
\end{lstlisting}
\begin{lstlisting}[captionpos=none,mathescape=true]
    /* Program 3 - Size of message */
if (h) then {
    out(Alice, "Hello");
} else {
    out(Alice, "");
}
\end{lstlisting}
\begin{lstlisting}[captionpos=none,mathescape=true]
    /* Program 4 - Time of message */
if (h) then {
    out(Alice, 42);
} else {
    sleep(100);
    out(Alice, 42);
}
\end{lstlisting}
As the above examples suggest, many convenient patterns in writing interactive programs are no longer secure when messages can be eavesdropped.

%
In this paper, we show that program level padding can be used for provably secure confidentiality against attackers observing the network trace. We do this by introducing SELENE, a Statically Enforced Language for Equivalence of Network Events. SELENE is a simple imperative programming language, that allows programmatic control over traffic shaping. We show that well-typed programs in SELENE satisfy timing-sensitive, progress-sensitive non-interference. We use a knowledge-based definition of non-interference \cite{askarov-myers-2011}
and show that an adversary learns no secrets by observing runs of well-typed programs. We assume that communication channels are partly observable. Namely, we assume that the presence of messages and the associated meta-information is publicly visible, while the contents of messages is only visible to trusted parties.

Our strategy for preventing traffic analysis attacks is to provide programmatic control over traffic shaping. We do this by splitting message sending into two distinct concepts: message allocation and message population.
To this end, SELENE uses two novel language primitives, \texttt{schedule} and \texttt{queue}, that respectively allocate a number of packets to be sent on a channel and add to a buffered output queue for a channel.
This simple strategy allows for utilising program level information to keep latency and bandwidth overheads low when compared with black-box padding. This property is particularly beneficial for resource constrained systems.
However, the strategy also comes with a downside, namely a restriction to when new traffic may be scheduled.

We present the formal semantics of the language in Section \ref{section:security_model_and_language}, but here present a few program examples possible in the language.

Consider a scenario where a doctor has asked a patient to take a home-test for an illness and to return the result. Depending on the result, the doctor may make a referral to a specialist clinic. Any message sent from the doctor to the clinic is publicly observable and plainly sending a referral will naturally leak that the patient returned a positive test result. However, if the doctor commits ahead of time to sending \textit{some} message to the clinic, regardless of the results of the test, the confidentiality of the patient's information can be protected.

\begin{lstlisting}[captionpos=none]
    /* Program 5 - Referral */
// Size of int
l_size = sizeof(0);

// Send to specialist in 300 time units
schedule(Clinic,l_size,300); 

// Get id and test result from patient
h_id = in(Patient);
h_is_positive = in(Patient);

if (h_is_positive) then {
    queue(Clinic,h_id);
} else {
    skip;
}
\end{lstlisting}
On line 6, the doctor schedules a send to the clinic in 300 time units. They await messages from the patient containing id number (line 9) and the test result (line 10), and if positive, the doctor queues a referral to the clinic (line 13). This strategy is somewhat optimistic as it may take more than 300 time units for the patient to send the result to the doctor. In this case, or in case the test result is negative, nothing will be queued to the clinic before the send occurs. If the queue is empty at the time of a scheduled send, dummy packets are sent instead. When, to whom, and how much the doctor sends is thereby made public, while \textit{what} the doctor sends is kept secret.

As a second example, we consider the password checker in Program 6.
\begin{lstlisting}[captionpos=none]
    /* Program 6 - Password checker */
string h_password;
int h_token;
l_size_ok = sizeof(h_token);
l_size_bad = sizeof("LOGIN FAILED");

schedule(Alice,max(l_size_ok,l_size_bad),100);
h_guess = in(Alice);
if (h_guess == h_password) then {
    queue(Alice,h_token);
} else {
    queue(Alice,"LOGIN FAILED");
}
\end{lstlisting}
The password checker stores a secret password and returns a token to be used as proof of authority upon receiving a successful guess. The password checker schedules bandwidth for sending either the token or a login failure message by using the maximum of the two sizes. By scheduling the response before the guess is received, the program does not leak whether a valid guess was received, let alone whether the guess was correct.

As a final example, we consider a small popularity poll. Alice wishes to know whether her opinion that dogs are better than cats is shared by a majority of people. She sets up a simple online voting service running Program 7 below:
\begin{lstlisting}[captionpos=none]
    /* Program 7 - Popinion */
/* Alice asks: Are cats or dogs better?
    Vote: Cats = 1, Dogs = -1 */
int h_my_vote;
l_tally = 0;
l_count = 0;

while (l_count < 10)
do {
    l_vote = in(Public);
    if (l_vote == -1 || l_vote == 1) then {
        l_tally = l_tally + l_vote;
    } else {
        skip;
    }
    l_count = l_count + 1;
}

// Size of the longest message
l_size = sizeof("Most disagree");
schedule(Alice, l_size, 100);

if (h_my_vote * l_tally > 0) then {
    queue(Alice, "Most agree");
} else if (h_my_vote * l_tally < 0) then {
    queue(Alice, "Most disagree");
} else {
    queue(Alice, "Tie");
}
\end{lstlisting}
The voting service stores Alice's secret choice in variable $\texttt{h\_my\_vote}$, tallies ten votes from a public channel, and schedules sending to Alice using the size of the longest message and time based on an estimate of what is needed for the branching. Inferring upper bounds on the time needed for queuing is orthogonal to the work in this paper and we opt for using simple estimates. Finally, the service computes whether a majority agrees or disagrees with Alice and sends a corresponding message.
We observe that no bandwidth is needed until ten votes have been received by the service. Since it cannot be determined statically when this occurs our approach reduces the traffic overhead induced compared with constant rate padding schemes as it allows scheduling of traffic on an as-needed basis, as long as the program context is public.

The main contributions of this paper are:
\begin{itemize}
    \item We spotlight the gap in the assumptions made in the language-based information flow literature for interactive programs and the channels available for real-world, online communication.
    \item We introduce SELENE, a language for using channels with observable traffic information in a principled and provably secure way, thereby recovering the strong security guarantees of language-based techniques.
    \item We introduce a novel model that combines program and runtime behaviour in a single small-step semantics, and give a knowledge-based security condition for timing-sensitive, progress-sensitive noninterference.
    \item We provide a progress-sensitive type system using both values of a fixed size type and values of a variable size type.
    \item We prove soundness of our type system, thereby obtaining a static guarantee that well-typed programs in SELENE do not leak via traffic patterns.
\end{itemize}


The remainder of this paper is structured as follows. In Section \ref{section:security_model_and_language} we specify the threat model and provide the syntax and semantics of SELENE. We define attacker knowledge and give a strong security condition against traffic analysis attacks in Section \ref{section:security_condition}. We present the security type system for SELENE and prove it sound in Section \ref{section:enforcement}. Finally, we discuss our work in Section \ref{section:discussion} and give related work in Section \ref{section:related_work}, before we conclude in Section \ref{section:conclusion}.


\section{Security model and language}
\label{section:security_model_and_language}
This section presents our model and the syntax and semantics of our language.

\subsection{Security model}
SELENE is an interactive, imperative language for single threaded, interactive programs with blocking receives. The language is largely standard, apart from our message sending primitives and command $\texttt{sizeof}$ for computing the size of a value. We assume a standard security lattice $\mathcal{L}$ of security levels $\ell$, with distinguished top and bottom elements, $\top$ and $\bot$, lattice ordering $\sqsubseteq$, and least upper bound operation $\sqcup$. Each variable has a fixed security level that does not change during execution.

As is standard in prior work on information flow control, we focus on confidentiality at the local node. Remote nodes trusted at some level $\ell$ are also trusted to appropriately protect information sent to them up to level $\ell$. We further assume that remote nodes are also running SELENE programs.
We model incoming traffic using lists. This modelling choice was shown equivalent to functional strategies for modelling deterministic, interactive programs by Clark and Hunt \cite{clark-hunt-2009}. To this end, we consider an input environment $\inenv$ mapping each channel to a (possibly empty) list of input packets and let program values be obtainable from a sequence of packets corresponding to the value.
For simplicity, we identify channels by their security level.

We observe that traffic analysis attacks exploit patterns in traffic to make inferences about the secret state of a system without requiring that the adversary can read the contents of packets. We therefore make the simplifying assumption that the contents of packets are sufficiently protected against adversaries, e.g. by using encryption, but allow the adversary to observe the presence, recipient, and time of packets. We assume that packets are of fixed size, thereby transforming the question of packet size into a question of packet count.

\subsection{Threat model}
We consider interactive, distributed programs that communicate with remote network nodes. We consider an active adversary who is trusted at a security level $\ladv$, who knows the program being run on the local node, and who knows initial secrets up to level $\ladv$. Additionally, the adversary eavesdrops on incoming and outgoing encrypted communication of the local node, observing packet presence, timing, and the remote communication party. Communication on a channel is encrypted corresponding to the security level of the channel, and the adversary can decrypt and read packets on channels up to security level $\ladv$. The objective of the adversary is to refine their knowledge on initial secrets.

\subsection{The language and program semantics}
\begin{figure}[!htb]
    \centering
\begin{framed}
\begin{align*}
    e \Coloneqq\;
    &n \mid s \mid x \mid e \oplus e\\
    c \Coloneqq\;
    &x = e \mid c; c \mid \texttt{skip} \mid \texttt{sleep} (e) \mid x = \texttt{sizeof}(e)\\
    \mid\; &\texttt{if } e \texttt{ then } c \texttt{ else } c \mid \texttt{while } e \texttt{ do } c\\
    \mid\; &x = \texttt{in}(\ell) \mid \texttt{schedule} (\ell,e,e) \mid \texttt{queue} (\ell,e)
\end{align*}
\end{framed}
    \caption{Syntax of the language}
    \label{fig:syntax_language}
\end{figure}

Figure \ref{fig:syntax_language} presents the syntax of our language. We explain the formal semantics and explain the nonstandard features. 

We use a big-step semantics for evaluating expressions and assume these take unit time. The rules are standard and are given in Fig. \ref{fig:semantics_expressions}.
\begin{figure}[!htb]
    \centering
\begin{framed}

\[
\inferrule {
    v \in \textit{Val}
} {
    \eval{v}{m}{v}
}
\hspace{3em}
\inferrule {
    m(x) = v
} {
    \eval{x}{m}{v}
}
\]

\[
\inferrule {
    \eval{e_1}{m}{v_1} \\
    \eval{e_2}{m}{v_2} \\
    v = v_1 \oplus v_2 \\
} {
    \eval{e_1 \oplus e_2}{m}{v}
}
\]

\end{framed}
    \caption{Semantics for evaluating expressions}
    \label{fig:semantics_expressions}
\end{figure}
We let $\oplus$ range over total operations on arithmetic expressions. The values of our language are integers $n$ and strings $s$. We let $\textit{Int}$ denote the set of integers and $\textit{String}$ denote the set of strings and let $\textit{Val} = \textit{Int}\, \uplus\, \textit{String}$. Programs are typed using fixed typing environment $\Gamma$. We write $\Gamma(x)=\ltype{\sigma}{\ell}$ to denote that variable $x$ has type $\sigma$ and security level $\ell$. The types of our language are $\inttype$ and $\stringtype{\ell}$, where $\ell$ is the security level of the size of the string. Input packets either contain (part of) an input value, or are dummy. We write $\xrect{v}^j_N$ to denote the $j$'th of $N$ packets encoding value $v$ and let $\xrect{\bullet}$ denote dummy packets.

For evaluating program commands $c$ we use a small-step semantics transition $\xangled{c,m,\inenv} \xrightarrow{\timestamp}_\alpha \xangled{c',m',\inenv'}$, where $m$ is a memory, $\inenv$ is an input environment, and $\alpha$ is a program event generated by the step. Program steps take place at a time $\timestamp$, however they do not increment time. We instead define a global semantics on top of the program semantics and let global steps increment time. We discuss the global semantics shortly.
Program events can be empty, denoted by $\epsilon$, or an assignment, enqueue, scheduling, or input event as given by the following grammar:
\begin{align*}
    \alpha \Coloneqq\;
    &\epsilon \mid \assignevent{x}{v} \mid \queueevent{\ell}{v} \mid  \scheduleevent{\ell}{n}{n} \mid \inputevent{\ell}{x}{v}
\end{align*}

Fig. \ref{fig:operational_semantics_program} presents the stepping rules of our program operational semantics.
\begin{figure*}[!htb]
    \centering
\begin{framed}

\[
\inferrule [\textsc{Assign}] {
    \xangled{e, m} \Downarrow v
} {
    \xangled{x = e,m,\inenv}
    \xrightarrow{\timestamp}_{\assignevent{x}{v}}
    \xangled{\texttt{stop},m[x \mapsto v],\inenv}
}
\hspace{3em}
\inferrule [\textsc{SizeOf}] {
    \xangled{e, m} \Downarrow v
    \\
    n = \left\lceil \frac{\textit{size}(v)}{\eta} \right\rceil
} {
    \xangled{x = \texttt{sizeof}(e),m,\inenv}
    \xrightarrow{\timestamp}_{\assignevent{x}{n}}
    \xangled{\texttt{stop},m[x \mapsto n],\inenv}
}
\]

\[
\inferrule [\textsc{Skip}] { } {
    \xangled{\texttt{skip},m,\inenv}
    \xrightarrow{\timestamp}_\epsilon
    \xangled{\texttt{stop},m,\inenv}
}
\hspace{1em}
\inferrule [\textsc{Seq-1}] {
    \xangled{c_1,m,\inenv}
    \xrightarrow{\timestamp}_\alpha
    \xangled{c'_1,m',\inenv'}
    \\
    c^\prime_1 \neq \texttt{stop}
} {
    \xangled{c_1;c_2,m,\inenv}
    \xrightarrow{\timestamp}_\alpha
    \xangled{c'_1;c_2,m',\inenv'}
}
\hspace{1em}
\inferrule [\textsc{Seq-2}] {
    \xangled{c_1,m,\inenv}
    \xrightarrow{\timestamp}_\alpha
    \xangled{\texttt{stop},m',\inenv'}
} {
    \xangled{c_1;c_2,m,\inenv}
    \xrightarrow{\timestamp}_\alpha
    \xangled{c_2,m',\inenv'}
}
\]

\[
\inferrule [\textsc{Sleep}] {
    \xangled{e, m} \Downarrow w \\
    w \geq 0 \\
    r = \timestamp + w
} {
    \xangled{\texttt{sleep}(e),m,\inenv}
    \xrightarrow{\timestamp}_\epsilon
    \xangled{\texttt{await}(r),m,\inenv}
}
\hspace{3em}
\inferrule [\textsc{Await}] {
    \timestamp \geq r
} {
    \xangled{\texttt{await}(r),m,\inenv}
    \xrightarrow{\timestamp}_\epsilon
    \xangled{\texttt{stop},m,\inenv}
}
\]

\[
\inferrule [\textsc{If-T}] {
    \eval{e}{m}{v} \\
    v \neq 0
} {
    \xangled{\texttt{if } e \texttt{ then } c_1 \texttt{ else } c_2,m,\inenv}
    \xrightarrow{\timestamp}_\epsilon
    \xangled{c_1,m,\inenv}
}
\hspace{3em}
\inferrule [\textsc{If-E}] {
    \eval{e}{m}{0}
} {
    \xangled{\texttt{if } e \texttt{ then } c_1 \texttt{ else } c_2,m,\inenv}
    \xrightarrow{\timestamp}_\epsilon
    \xangled{c_2,m,\inenv}
}
\]

\[
\inferrule [\textsc{While}] { } {
    \xangled{\texttt{while } e \texttt{ do } c,m,\inenv}
    \xrightarrow{\timestamp}_\epsilon
    \xangled{\texttt{if } e \texttt{ then } c;\texttt{while } e \texttt{ do } c \texttt{ else skip},m,\inenv}
}
\]

\[
\inferrule [\textsc{In}] {
    A \in \{\textit{Int},\textit{String} \} \\
    m(x) \in A \\
    \inenv(\ell) = \vec{p} \\
    (v,\vec{q}) = \textsf{choose}(\vec{p},A,\timestamp,[])
} {
    \xangled{x = \texttt{in}(\ell),m,\inenv}
    \xrightarrow{\timestamp}_{\inputevent{\ell}{x}{v}}
    \xangled{\texttt{stop},m[x \mapsto v],\inenv[\ell \mapsto \vec{q}]}
}
\]

\[
\inferrule [\textsc{Schedule}] {
    \eval{e_1}{m}{n} \\
    \eval{e_2}{m}{w} \\
    w \geq 0 \\
    t = \timestamp + w
} {
    \xangled{\texttt{schedule}(\ell,e_1,e_2),m,\inenv}
    \xrightarrow{\timestamp}_{\scheduleevent{\ell}{n}{t}}
    \xangled{\texttt{stop},m,\inenv}
}
\hspace{3em}
\inferrule [\textsc{Queue}] {
    \eval{e}{m}{v}
} {
    \xangled{\texttt{queue}(\ell,e),m,\inenv}
        \xrightarrow{\timestamp}_{\queueevent{\ell}{v}}
        \xangled{\texttt{stop},m,\inenv}
}
\]

\end{framed}
    \caption{Local operational semantics}
    \label{fig:operational_semantics_program}
\end{figure*}

\subsubsection*{SizeOf}
Command \texttt{sizeof} evaluates an expression to obtain a value and returns the number of packets needed to store that value. This is useful as the language requires the programmer to explicitly schedule the number of packets they wish to send. We assume a fixed packet size $\eta$, and assume that all integers are of fixed size, and that the size of a string is dependent on the length of the string.

\subsubsection*{In}
The transitions of our small-step semantics for program commands are parametric in timestamps $\timestamp$, allowing us to model blocking by conditioning transitions on $\timestamp$. We model network input using primitive $x = \texttt{in}(\ell)$. To preserve the type of variable $x$, the input primitive determines whether $m(x)$ is an integer or a string value, captured by set $A$ in rule \textsc{In}, and uses this as argument for auxiliary function \textsf{choose} (Fig. \ref{fig:choose_function}), along with the packet sequence for channel $\ell$ and timestamp $\timestamp$. Function \textsf{choose} is a partial function, modelling the potential for blocking. The \textsf{choose} function adds packets of the appropriate type to an accumulator used for decoding an input value, and steps over packets of other type, discarding any dummy packets. It returns a decoded value and the remaining packet sequence for the channel if successful.

\begin{figure*}[!htb]
    \centering
    \[
    \textsf{choose}(\vec{p},A,t,\textit{acc}) \triangleq
        \begin{cases}
            (v,\vec{p})
                &\text{if } \textit{acc} = \xrect{v}^1_N \Colon \ldots \Colon \xrect{v}^N_N \\
            (v_r, \vec{r})
                &\text{if } \vec{p} = (t',\xrect{v}^j_N) \cdot \vec{q}
                \text{ s.t. } v \in A
                \text{ and } t' \leq t
                \text{ and } \textsf{choose}(\vec{q},A,t,\textit{acc} \Colon \xrect{v}^j_N) = (v_r,\vec{r}) \\
            (v_r,(t',\xrect{v}^j_N) \cdot \vec{r})
                &\text{if } \vec{p} = (t',\xrect{v}^j_N) \cdot \vec{q}
                \text{ s.t. } v \notin A
                \text{ and } t' \leq t
                \text{ and } \textsf{choose}(\vec{q},A,t,\textit{acc}) = (v_r,\vec{r}) \\
            (v_r,\vec{r})
                &\text{if } \vec{p} = (t',\xrect{\bullet}) \cdot \vec{q}
                \text{ s.t. } t' \leq t
                \text{ and } \textsf{choose}(\vec{q},A,t,\textit{acc}) = (v_r,\vec{r})
        \end{cases}
    \]
    \caption{Choose function}
    \label{fig:choose_function}
\end{figure*}
We let the packets be annotated with timestamps and require that all packets corresponding to a value have been received before the value can be obtained. That is, the timestamp of the final packet must be at or before the timestamp in the transition of the \texttt{in} command. If no value can be retrieved, the program blocks and cannot step.

We model internal input (e.g. reading files) only abstractly, by considering them bound in program variables.

\subsubsection*{Schedule and queue}
The \texttt{schedule} command takes three arguments; a channel, a number of packets to be sent, and a delay before sending the packets. This issues a request to the runtime system. We describe the runtime system shortly. Command \texttt{queue} takes a channel and an expression as arguments and evaluates the expression to obtain a value. It then instructs the runtime system to add the value to a buffered output queue associated with the channel.

\subsubsection*{Internal commands}
Commands \texttt{await} and \texttt{stop} are only used internally and are therefore not part of the language syntax. Command \texttt{await} is reached from command \texttt{sleep} and blocks for a specified duration of time. Command \texttt{stop} denotes a final program configuration that cannot step any further.

\subsection{The runtime and the global semantics}
A key feature of our model is the global configuration modelling the language runtime. The runtime maintains the output queues in output environment $\outenv$ and processes the packet schedule $\schedule$. We present a small-step semantics for the global transitions in Fig. \ref{fig:operational_semantics_global}.
\begin{figure*}[!htb]
    \centering
\begin{framed}

\[
\inferrule [\textsc{G-Step}] {
    \lconfig \xrightarrow{\timestamp}_{\alpha} \lconfig' \\
    (\outenv',\schedule') = \textsf{upd}(\outenv,\schedule,\alpha) \\
    (\beta,\outenv^\dprime) =
        {\begin{cases}
            (\epsilon,\outenv')
                &\text{if } \timestamp \notin \textsf{dom}(\schedule')\\
            \textsf{send}(\outenv',\ell)
                &\text{if } \schedule'(\timestamp) = \ell
        \end{cases}}
} {
    \xAngled{\lconfig,\outenv,\schedule,\timestamp}
    \rightarrowdbl_{(\timestamp:\alpha,\beta)}
    \xAngled{\lconfig',\outenv^\dprime,\schedule^\prime,\timestamp+1}
}
\]

\[
\inferrule [\textsc{G-Block}] {
    \not\exists \lconfig' : \lconfig \xrightarrow{\timestamp}_{\alpha} \lconfig' \\
    \lconfig = \xangled{c,m,\inenv} \\
    c \neq \texttt{stop} \\
    (\beta,\outenv') =
        {\begin{cases}
            (\epsilon,\outenv)
                &\text{if } \timestamp \notin \textsf{dom}(\schedule)\\
            \textsf{send}(\outenv,\ell)
                &\text{if } \schedule(\timestamp) = \ell
        \end{cases}}
} {
    \xAngled{\lconfig,\outenv,\schedule,\timestamp}
    \rightarrowdbl_{(\timestamp:\epsilon,\beta)}
    \xAngled{\lconfig,\outenv',\schedule,\timestamp+1}
}
\]

\[
\inferrule [\textsc{G-Stop}] {
    \lconfig = \xangled{\texttt{stop},m,\inenv} \\
    \exists \timestamp' \in \textsf{dom}(\schedule) : \timestamp' \geq \timestamp \\
    (\beta,\outenv') =
        {\begin{cases}
            (\epsilon,\outenv)
                &\text{if } \timestamp \notin \textsf{dom}(\schedule)\\
            \textsf{send}(\outenv,\ell)
                &\text{if } \schedule(\timestamp) = \ell
        \end{cases}}
} {
    \xAngled{\lconfig,\outenv,\schedule,\timestamp}
    \rightarrowdbl_{(\timestamp:\epsilon,\beta)}
    \xAngled{\lconfig,\outenv',\schedule,\timestamp+1}
}
\]

\end{framed}
    \caption{Global operational semantics}
    \label{fig:operational_semantics_global}
\end{figure*}

We explain the interaction between the program configuration and global configuration in more detail.
We let the global configuration contain a program configuration and require that the program steps whenever possible. For the sake of brevity, we let $\lconfig$ denote a program configuration $\langle c, m ,\inenv \rangle$.
Program steps are done by rule \textsc{G-Step} and emit a possibly empty event $\alpha$. We let the program communicate updates to the runtime through schedule and queue events. To this end, we apply update function \textsf{upd} (Fig. \ref{fig:runtime_function}) to the schedule and output environment using event $\alpha$.
\begin{figure}[!htb]
    \centering
\begin{align*}
    \textsf{split}(v) &\triangleq \xrect{v}^1_N \cdot \ldots \cdot \xrect{v}^N_N
    \qquad \text{where } N = \left\lceil \frac{\textit{size}(v)}{\eta} \right\rceil
\end{align*}
\begin{align*}
    \textsf{rsv}(\schedule,\ell,n,t) &\triangleq
    \begin{cases}
        \schedule_r
            &\text{if } n > 0
            \text{ and } t \not\in \textsf{dom}(\schedule) \text{ and}\\
            &\textsf{rsv}(\schedule[t \mapsto \ell],\ell,n-1,t+1) = \schedule_r\\
        \schedule_r
            &\text{if } n > 0
            \text{ and } t \in \textsf{dom}(\schedule) \text{ and} \\
            &\textsf{rsv}(\schedule,\ell,n,t+1) = \schedule_r\\
        \schedule
            &\text{if } n \leq 0
    \end{cases}
    \\
    \textsf{upd}(\outenv,\schedule,\alpha) &\triangleq
    \begin{cases}
         (\outenv',\schedule)
            &\text{if } \alpha = \queueevent{\ell}{v}
            \text{, } \textsf{split}(v) = \vec{p},\\
            &\outenv(\ell) = \packetqueue,
            \text{and } \outenv[\ell \mapsto (\packetqueue \cdot \vec{p})] = \outenv'\\
        (\outenv,\schedule')
            &\text{if } \alpha = \scheduleevent{\ell}{n}{t} \text{ and}\\
            &\textsf{rsv}(\schedule,\ell,n,t)=\schedule'\\
        (\outenv,\schedule)
            &\text{otherwise}
    \end{cases}
\end{align*}
\caption{Runtime function}
    \label{fig:runtime_function}
\end{figure}
If the program step emits a queue event $\queueevent{\ell}{v}$, value $v$ is split into a number of packets based on the size of the value, and the packets are added to the buffered output queue for channel $\ell$. If the program step emits a schedule event $\scheduleevent{\ell}{n}{t}$, we use function $\textsf{rsv}$ to reserve time in the schedule for a number of packets on a channel by recursively adding to $\textsf{dom}(\schedule)$. 
We assume that the schedule is never full, i.e. the function will terminate having scheduled all $n$ packets. We make the simplifying assumption that at most one packet can be sent in any single step and formally model the schedule as a partial function from timestamps to channels.

To model packets being sent, we extend the grammar for events with runtime events $\beta$. The runtime emits an empty event if the schedule is undefined for the current timestamp, otherwise we use function \textsf{send} defined below to obtain an event corresponding to the first packet in the scheduled channel's output queue, and an updated output environment. If no packets are queued on the channel, an empty dummy packet is generated and sent.
\[
    \textsf{send}(\outenv,\ell) \triangleq
    \begin{cases}
        (\outputevent{\ell}{p},\outenv[\ell \mapsto \packetqueue])
            &\text{if } \outenv(\ell) = p \cdot \vec{q}\\
        (\outputevent{\ell}{\xrect{\bullet}},\outenv)
            &\text{if } \outenv(\ell) = []
    \end{cases}
\]
To combine program generated events $\alpha$ with runtime generated events $\beta$ we let global events $\gamma$ be a triple $(\timestamp: \alpha, \beta)$, where $\timestamp$ is the timestamp of the event. The observations an attacker makes on a run of a program are given by a trace of global events, each containing the timestamp of the step, leading to a timing-sensitive model. While a network attacker does not observe program events $\alpha$, maintaining them in global events is nevertheless useful, as it allows us to more easily reason about the exact state of a run. In Section \ref{section:security_condition}, we define our security condition in terms of an attacker that does not observe program events, i.e., that observes all program events as the empty event $\epsilon$.

We extend the grammar as follows:
\begin{align*}
    \beta \Coloneqq\;
    &\epsilon \mid \outputevent{\ell}{p}\\
    \gamma \Coloneqq\;
    &(\timestamp: \alpha,\beta)
\end{align*}

The global configuration maintains clock $\timestamp$ that is incremented for each step. If a program configuration is blocking, that is, if it cannot take a step at the current timestamp, but has not stopped with command $\texttt{stop}$, the global configuration steps by \textsc{G-Block}, processing the runtime and incrementing the clock. Finally, \textsc{G-Stop} allows the global configuration to continue processing the runtime after the program configuration has reached command $\texttt{stop}$, provided there are scheduled packets left to process.

\section{Security condition}
\label{section:security_condition}
In this section we present the security condition for timing-sensitive, progress-sensitive noninterference.

We define our security condition using the knowledge-based approach \cite{askarov-chong-2012}. 
The insight of this approach is to consider what an attacker observes during the execution of a program and define knowledge as the set of initial states that are consistent with seeing the execution up to this point. The security condition is then defined as a bound on how much the knowledge is allowed to change for each step of the execution. In this paper we do not consider declassification and we therefore require that attacker knowledge does not change with new observations.

\subsection{Auxiliary definitions}
We define attacker knowledge and timing-sensitive, progress-sensitive noninterference in terms of an equivalence relation on program configurations and the attacker observable trace emitted from a run. We give these auxiliary definitions before proceeding to define the security condition.

To denote that two memories are equivalent up to $\ladv$ we write $m \ladvequiv m'$ (Definition \ref{definition:memory_equivalence}).
\begin{definition}[Memory equivalence up to level] \label{definition:memory_equivalence}
Two memories $m$ and $m'$ are equivalent up to level $\ladv$, written $m \ladvequiv m'$, if for all $x \in \textsf{dom}(\Gamma)$ both the following hold:
\begin{enumerate}
    \item $\Gamma(x) = \ltype{\sigma}{\ell} \land \ell \sqsubseteq \ladv \implies m(x) = m'(x)$
    \item $\Gamma(x) = \ltype{\stringtype{\ell'}}{\ell} \land \ell' \sqsubseteq \ladv \implies
    \textit{size}(m(x)) = \textit{size}(m'(x))$
\end{enumerate}
\end{definition}
This definition captures that the values of attacker observable variables must have the same value, and that the size of the value of variables must be the same if the size is attacker observable.

We overload the notation and write $\inenv \ladvequiv \inenv'$ to denote that two input environments are equivalent up to $\ladv$ (Definition \ref{definition:input_equivalence}). The definition requires equality of incoming packets on attacker observable channels, and uses relation $\approx^\textit{net}_{\ladv}$ for high channels, requiring that these receive packets at the same timestamps. This captures an attacker that can observe the presence of incoming packets on all channels and when they arrive, but who cannot read the contents of packets on high channels. We assume that incoming packets are sent by other SELENE programs and hence are of fixed size.

\begin{definition}[Input environment equivalence up to level]  \label{definition:input_equivalence}
Two input environments $\inenv$ and $\inenv'$ are equivalent up to level $\ladv$, written $\inenv \ladvequiv \inenv'$, if
\[
\inferrule {
    \ell \sqsubseteq \ladv \implies \inenv_1(\ell) = \inenv_2(\ell) \\
    \ell \not\sqsubseteq \ladv \implies \inenv_1(\ell) \ladvequivnet \inenv_2(\ell)
} {
    \inenv_1 \ladvequiv \inenv_2
}
\]
where $\approx^\textit{net}_{\ladv}$ is defined by
\[
\inferrule {
} {
    [] \ladvequivnet []
}
\hspace{1em}
\inferrule {
    \vec{p} \ladvequivnet \vec{q}
} {
    (t,p_1) \cdot \vec{p} \ladvequivnet (t,p_2) \cdot \vec{q}
}
\]
\end{definition}

We lift equivalences to program configurations in a straight forward way in Definition \ref{definition:lconfig_equivalence}.

\begin{definition}[Program configuration equivalence up to level] \label{definition:lconfig_equivalence}
Two program configurations $\xangled{c_1,m_1,\inenv_1}$ and $\xangled{c_2,m_2,\inenv_2}$ are equivalent up to level $\ladv$, written
\[
    \xangled{c_1,m_1,\inenv_1} \ladvequiv \xangled{c_2,m_2,\inenv_2}
\]
if it holds that $c_1 = c_2$, $m_1 \ladvequiv m_2$, and $\inenv_1 \ladvequiv \inenv_2$.
\end{definition}

We overload the notation even further and write $\outenv \ladvequiv \outenv'$ to denote that two output environments are equivalent up to $\ladv$ (Definition \ref{definition:output_equivalence}).

\begin{definition}[Output environment equivalence up to level]  \label{definition:output_equivalence}
Two output environments $\outenv_1$ and $\outenv_2$ are equivalent up to level $\ladv$, written $\outenv_1 \ladvequiv \outenv_2$, if
\[
\inferrule {
    \ell \sqsubseteq \ladv \implies \outenv_1(\ell) = \outenv_2(\ell)
} {
    \outenv_1 \ladvequiv \outenv_2
}
\]
\end{definition}

Next, we define runtime event projections. Runtime event projection captures the observable parts of output events emitted by the runtime. We write $\projevent{\beta}{\ladv}$ to denote the projection of runtime event $\beta$ to level $\ladv$ (Definition \ref{definition:runtime_event_projection}). We introduce a new event capturing the sending of packets with unobservable content. We extend the grammar for runtime events as follows:
\begin{align*}
    \beta \Coloneqq\;
    &\ldots \mid \outputevent{\ell}{-}
\end{align*}
Runtime event $\outputevent{\ell}{p}$ projects to $\outputevent{\ell}{-}$ if the level of the channel does not flow to the level being projected to. This captures the assumption that the contents of packets can be securely hidden by cryptography, while the presence of packets and their recipient remain visible.

\begin{definition}[Runtime event projection] \label{definition:runtime_event_projection}
The projection of runtime event $\beta$ to level $\ladv$, written $\projevent{\beta}{\ladv}$, is defined as
\begin{align*}
    \projevent{\epsilon}{\ladv} &= \epsilon
    \\
    \projevent{\outputevent{\ell}{p}}{\ladv} &=
    \begin{cases}
        \outputevent{\ell}{p}
            &\text{if } \ell \sqsubseteq \ladv\\
        \outputevent{\ell}{-}
            &\text{if } \ell \not\sqsubseteq \ladv
    \end{cases}
\end{align*}
\end{definition}

We write $\filtertrace{\tau}{\ladv}$ to denote the filtering of trace $\tau$ to what is visible at level $\ladv$ (Definition \ref{definition:trace_filtering}). We use this to restrict attacker knowledge to the steps where observable events are emitted. This allows us to consider secure programs such as $\texttt{if h then sleep(10) else skip}$, as no output occurs after branching on the secret. We let trace filtering fix the empty event $\epsilon$ as the program event component, thereby removing all program events emitted. This captures a network attacker, that only obtains new information by observing packets being sent.

\begin{definition}[Trace filtering] \label{definition:trace_filtering}
The filtering of a trace $\tau$ to level $\ladv$, written $\filtertrace{\tau}{\ladv}$, is defined as
\begin{align*}
    \filtertrace{\epsilon}{\ladv} &= \epsilon\\
    \filtertrace{(\tau' \cdot (\timestamp: \alpha, \beta))}{\ladv} &= \\
    &\hspace{-5em}
    \begin{cases}
        \filtertrace{\tau'}{\ladv} \cdot (\timestamp: \epsilon, \projevent{\beta}{\ladv})
            &\text{if } \projevent{\beta}{\ladv} \neq \epsilon\\
        \filtertrace{\tau'}{\ladv}
            &\text{otherwise}
    \end{cases}
\end{align*}
\end{definition}

As two final building blocks, we let $\outenv_\textit{init}$ denote the initially empty output environment and let $\schedule_\textit{init}$ denote the initially empty schedule. That is, 
\begin{align*}
    \forall \ell \in \mathcal{L}: \outenv_\textit{init}(\ell) &= []\\
    \textsf{dom}(\schedule_\textit{init}) &= \emptyset
\end{align*}

\subsection{Knowledge and noninterference}
Using the above we define the knowledge of an attacker at level $\ladv$ after observing trace $\tau$. This definition follows the style of other knowledge-based security conditions \cite{askarov2007gradual}, and intuitively states that an attacker may not refine their knowledge by observing new events.

\begin{definition}[Attacker knowledge] \label{definition:attacker_knowledge}
Given a program configuration $\lconfig$, such that $\xAngled{\lconfig,\outenv_\textit{init},\schedule_\textit{init},0}
\rightarrowdbl^*_{\tau}
\xAngled{\lconfig',\outenv',\schedule',\timestamp'}$, the attacker knowledge at level $\ladv$ is the set of program configurations $\lconfig_2$, that are consistent with observations at that level:
\begin{align*}
    &k(\lconfig,\tau,\ladv) \triangleq \\
    &\quad
    \{ \lconfig_2 \mid \lconfig \ladvequiv \lconfig_2\; \land\\
    &\qquad
    \xAngled{\lconfig_2,\outenv_\textit{init},\schedule_\textit{init},0}
    \rightarrowdbl^*_{\tau_2}
    \xAngled{\lconfig'_2,\outenv'_2,\schedule'_2,\timestamp'_2}\; \land\\
    &\qquad
    (\filtertrace{\tau}{\ladv}) = (\filtertrace{\tau_2}{\ladv}) \}
\end{align*}
\end{definition}

Using the definition of attacker knowledge we define timing-sensitive, progress-sensitive noninterference.

\begin{definition}[Timing-sensitive, progress-sensitive noninterference] \label{definition:timing-sensitive_noninterference}
Given program configuration $\lconfig$ such that
\[
    \xAngled{\lconfig,\outenv_\textit{init},\schedule_\textit{init},0}
    \rightarrowdbl^*_{\tau \cdot \gamma}
    \xAngled{\lconfig',\outenv',\schedule',\timestamp'}
\]
the run satisfies timing-sensitive, progress-sensitive noninterference if for all $\ladv$ it holds that
\[
    k(\lconfig,\tau \cdot \gamma,\ladv) \supseteq k(\lconfig,\tau,\ladv)
\]
\end{definition}
This definition states that memories and input environments considered possible before observing global event $\gamma$ are also considered possible after observing $\gamma$, capturing that the adversary learns nothing by observing the event.
To demonstrate the security condition, we rewrite Program 3 from Section \ref{section:introduction} in the syntax of SELENE. We consider one run where secret variable $\texttt{h}$ is set to 1 and another where it is set to 0, and assume that $n+1$ packets are needed to send a string of length $n$.
\begin{lstlisting}[captionpos=none]
    /* Program 3b */
if (h) then {
    queue(Alice, "Hello");
    size = sizeof("Hello");
    schedule(Alice,size,0);
} else {
    queue(Alice, "");
    size = sizeof("");
    schedule(Alice,size,0);
}
\end{lstlisting}
In the first run, 6 packets are scheduled and sent on channel $\texttt{Alice}$ in order to send the string. In the second run, only a single packet is scheduled and sent. As the presence of every packet is observable to the attacker, they can distinguish between the two runs when a second packet is sent, hence violating Definition \ref{definition:timing-sensitive_noninterference}.

\section{Enforcement} 
\label{section:enforcement}
In this section we present the security type system for SELENE and prove that all runs of well-typed programs satisfy timing-sensitive, progress-sensitive noninterference.
Despite considering an attacker that observes only network events, timing-sensitivity makes the attacker quite strong. We settle for a rather restrictive type system that is secure against a stronger, internal attacker and we use this to show security for a network attacker.
Previous work on noninterference for interactive programs by O'Neill et al. \cite{oneill-clarkson-chong-2006} achieves a more permissive type system by assuming a timing-insensitive attacker, disallowing high-loops, and assuming that new input is always available, thereby ruling out high-divergence of programs. Unfortunately, timing-insensitive attacker models are insufficient for external attackers, such as the network attacker we consider, as we cannot restrict an attacker's access to timing channels.

As our security condition is timing-sensitive and progress-sensitive, our typing judgements are progress-sensitive. They are of the form
\[
    \Gamma, \pc \vdash c : \pc'
\]
where $\pc$ is the program counter before typing the command and $\pc'$ is the program counter after. As we do not consider $\pc$-declassification in this paper, the program counter never goes down. This leads to so-called $\pc$-creep, which significantly restricts the programs that can be written in the language. We leave it to future work to explore language primitives for mitigating this and to investigate the security impact of $\pc$-declassification on traffic analysis attacks.

\subsection{Type system}
The definitions of memory equivalence and event projection in Section \ref{section:security_condition} implicitly require a well-formedness condition on security types of variables in $\Gamma$. We now formally state this condition and assume for the rest of the paper that all variables in $\Gamma$ have well-formed security types.
\[
\inferrule { } {
    \vdash_\textsf{wf} \ltype{\inttype}{\ell}
}
\hspace{3em}
\inferrule { 
    \ell' \sqsubseteq \ell
} {
    \vdash_\textsf{wf} \ltype{\stringtype{\ell'}}{\ell}
}
\]
The intuition behind the well-formedness condition is that knowing a value implies knowing its size, but not the other way around. Next, we define subtyping relation $<:$. The relation is straight forward, using lattice ordering $\sqsubseteq$ on size levels as a condition on strings.
\[
\inferrule { } {
    \inttype <: \inttype
}
\hspace{3em}
\inferrule {
    \ell_1 \sqsubseteq \ell_2
} {
    \stringtype{\ell_1} <: \stringtype{\ell_2}
}
\]
We write $\sigma \nearrow \ell$ to denote raising type $\sigma$ to at least level $\ell$. This is used to account for $\pc$-taint when assigning strings, to prevent string size from leaking secrets. This allows us to concisely write the conditions for the typing rules.
\[
    \sigma \nearrow \ell \triangleq
    \begin{cases}
        \inttype
            &\text{if } \sigma = \inttype\\
        \stringtype{(\ell \sqcup \ell')}
            &\text{if } \sigma = \stringtype{\ell'}
    \end{cases}
\]

The type system for expressions is presented in Fig. \ref{fig:type_system_expressions}. The rules are standard, except for the rule for string expressions. This rule follows from the well-formedness condition on security types, and intuitively as the string appears in the program text, hence the size of the string is public.

\begin{figure}[!htb]
    \centering
\begin{framed}

\[
\inferrule {
    n \in \textit{Int}
} {
    \Gamma \vdash n : \ltype{\inttype}{\bot}
}
\hspace{1em}
\inferrule {
    s \in \textit{String}
} {
    \Gamma \vdash s : \ltype{\stringtype{\bot}}{\bot}
}
\hspace{1em}
\inferrule { } {
    \Gamma \vdash x : \Gamma(x)
}
\]

\[
\inferrule {
    \Gamma \vdash e_1 : \ltype{\inttype}{\ell_1} \\
    \Gamma \vdash e_2 : \ltype{\inttype}{\ell_2}
} {
    \Gamma \vdash e_1 \oplus e_2 : \ltype{\inttype}{\ell_1 \sqcup \ell_2}
}
\]
%

\end{framed}
    \caption{Type system for expressions}
    \label{fig:type_system_expressions}
\end{figure}

We present our type system for commands in Fig. \ref{fig:type_system_commands} and explain the nonstandard rules.
\begin{figure*}[!htb]
    \centering
\begin{framed}

\[
\inferrule[\textsc{T-Assign}] {
    \Gamma \vdash e : \ltype{\sigma_e}{\ell_e} \\
    \sigma_e \nearrow \pc <: \sigma_x \\\\
    \Gamma(x) = \ltype{\sigma_x}{\ell_x} \\
    \ell_e \sqcup \pc \sqsubseteq \ell_x
} {
    \Gamma, \pc \vdash x = e : \pc
}
\hspace{5em}
\inferrule[\textsc{T-Skip}] { } {
    \Gamma, \pc \vdash \texttt{skip} : \pc
}
\hspace{5em}
\inferrule[\textsc{T-Sleep}] {
    \Gamma \vdash e : \ltype{\inttype}{\ell}
} {
    \Gamma,\pc \vdash \texttt{sleep}(e) : \pc \sqcup \ell
}
\]

\[
\inferrule[\textsc{T-SizeOf}] {
    \Gamma \vdash x : \ltype{\inttype}{\ell_x} \\
    \Gamma \vdash e : \ltype{\sigma_e}{\ell_e} \\
    \pc \sqsubseteq \ell_x \\
    \sigma_e = \stringtype{\ell'} \implies \ell' \sqsubseteq \ell_x
} {
    \Gamma, \pc \vdash x = \texttt{sizeof}(e) : \pc
}
\]

\[
\inferrule[\textsc{T-Await}] { } {
    \Gamma,\pc \vdash \texttt{await}(r) : \pc
}
\hspace{5em}
\inferrule[\textsc{T-If}] {
    \Gamma \vdash e : \ltype{\inttype}{\ell} \\
    \Gamma, \pc \sqcup \ell \vdash c_1 : \pc^\prime\\
    \Gamma, \pc \sqcup \ell \vdash c_2 : \pc^\dprime
} {
    \Gamma, \pc \vdash \texttt{if } e \texttt{ then } c_1 \texttt{ else } c_2 : \pc^\prime \sqcup \pc^\dprime
}
\]

\[
\inferrule[\textsc{T-Seq}] {
    \Gamma, \pc \vdash c_1 : \pc^\prime \\
    \Gamma, \pc^\prime \vdash c_2 : \pc^\dprime
} {
    \Gamma, \pc \vdash c_1;c_2 : \pc^\dprime
}
\hspace{5em}
\inferrule[\textsc{T-While}] {
    \Gamma \vdash e : \ltype{\inttype}{\ell} \\
    \Gamma, \pc \sqcup \ell \vdash c : \pc^\prime
} {
    \Gamma, \pc \vdash \texttt{while } e \texttt{ do } c : \pc^\prime
}
\]

\[
\inferrule[\textsc{T-In}] {
    \Gamma \vdash x : \ltype{\sigma_x}{\ell_x} \\
    \pc \sqsubseteq \ell \\
    \sigma_x \nearrow \ell <: \sigma_x \\
    \ell \sqsubseteq \ell_x
} {
    \Gamma, \pc \vdash x = \texttt{in}(\ell) : \ell
}
\]

\[
\inferrule[\textsc{T-Schedule}] {
    \pc = \bot \\
    \Gamma \vdash e_1 : \ltype{\inttype}{\bot} \\
    \Gamma \vdash e_2 : \ltype{\inttype}{\bot}
} {
    \Gamma, \pc \vdash \texttt{schedule}(\ell,e_1,e_2) : \pc
}
\hspace{5em}
\inferrule[\textsc{T-Queue}] {
    \Gamma \vdash e : \ltype{\sigma_e}{\ell_e} \\
    \ell_e \sqcup \pc \sqsubseteq \ell
} {
    \Gamma, \pc \vdash \texttt{queue}(\ell,e) : \pc
}
\]

\end{framed}
    \caption{Type system for commands}
    \label{fig:type_system_commands}
\end{figure*}

\subsubsection*{SizeOf}
Rule $\textsc{T-SizeOf}$ expresses that the size of an integer value may be assigned to a variable conditioned only by $\pc$. This is intuitively safe as integers have fixed size. The size of a string value may be assigned to a variable if the variable is at least as secret as the least upper bound of $\pc$ and the size level of the string.

\subsubsection*{In}
Rule $\textsc{T-In}$ is similar to input rules in previous work. We require that the level of $\pc$ flows to the level of the channel $\ell$. This is to preserve low equivalence of the input environment during steps under high $\pc$. As a non-standard condition, the rule uses type raising and the subtyping relation to require $\sigma_x \nearrow \ell <: \sigma_x$. For $\sigma_x = \inttype$, this condition is trivially satisfied by the definitions. For $\sigma_x = \stringtype{\ell'}$, this condition corresponds to the condition $\ell \sqsubseteq \ell'$. Intuitively, we consider both the size and the value of a received strings to be as secret as the level of the channel.

\subsubsection*{Schedule and queue}
Rule $\textsc{T-Schedule}$ restricts schedule commands to public $\pc$ and restricts the integer arguments to also be public. These conditions are natural, as we assume that traffic is publicly observable. As a consequence of progress-sensitive typing, a schedule command cannot occur after the $\pc$ has been tainted. The queuing of messages is by rule $\textsc{T-Queue}$ less restrictive, and is akin to rules for sending in previous information flow literature.

We note in particular that Programs 5, 6, and 7 from Section \ref{section:introduction} are typeable by the typing rules, while the rewritten Program 3b from Section \ref{section:security_condition} is not as it performs scheduling after branching on a high variable.

\subsection{Program configuration}
We show soundness of our security type system in a number of steps. We show that the type system of SELENE is secure against a stronger, internal attacker and show that this implies security against an external attacker. This is intuitively safe as the external attacker has weaker observational power.

We begin by defining well-formedness conditions on memories (Definition \ref{definition:memory_well-formedness_wrt_typing_environment}) and program configurations (Definition \ref{definition:lconfig_well-formedness_wrt_typing_environment}). These conditions are standard. We define memory $m$ to be well-formed with respect to typing environment $\Gamma$ in the straight forward way.

\begin{definition}[Well-formedness of memory w.r.t. a typing environment] \label{definition:memory_well-formedness_wrt_typing_environment}
Given a memory $m$ and a typing environment $\Gamma$, we say that $m$ is well-formed w.r.t. $\Gamma$ if for all $x \in \textsf{dom}(\Gamma)$ we have
\begin{enumerate}
    \item[(1)] $m(x) \in \textit{Int} \implies \Gamma(x) = \ltype{\inttype}{\ell}$
    \item[(2)] $m(x) \in \textit{String} \implies \Gamma(x) = \ltype{\stringtype{\ell'}}{\ell}$
\end{enumerate}
\end{definition}

We define program configuration $\xangled{c,m,\inenv}$ to be well formed with respect to a typing environment $\Gamma$ and program counters $\pc, \pc'$ if $c$ is \texttt{stop} of if $c$ is typable, and if $m$ is well-formed with respect to $\Gamma$.

\begin{definition}[Well-formedness of program configurations] \label{definition:lconfig_well-formedness_wrt_typing_environment}
We say that program configuration $\xangled{c,m,\inenv}$ is well-formed w.r.t. a typing environment $\Gamma$ and levels $\pc, \pc'$ when both the following hold:
\begin{enumerate}
    \item[(1)] either $c$ is $\texttt{stop}$ or the program is well-typed, i.e., $\Gamma,\pc \vdash c : \pc'$
    \item[(2)] $m$ is well-formed w.r.t. $\Gamma$
\end{enumerate}
\end{definition}

Steps of the program preserve well-formedness by Lemma \ref{lemma:preservation_well-formedness}. The proof can be found in
\ifdefined\istechnicalreport
the Appendix.
\else
the accompanying technical report.
\fi
\begin{restatable}[Preservation of well-formedness]{lemma}{preservationwellformedness}
\label{lemma:preservation_well-formedness}
Let $\Gamma$ be a typing environment, $\pc,\pc'$ be two levels, and $\xangled{c,m,\inenv}$ be a program configuration, such that the $\xangled{c,m,\inenv}$ is well-formed w.r.t. $\Gamma$, $\pc$, and $\pc'$. Suppose this configuration takes a step
\[
    \xangled{c,m,\inenv}
    \xrightarrow{\timestamp}_{\alpha}
    \xangled{c',m',\inenv'}
\]
Then there exists $\pc^\dprime$ such that $\pc \sqsubseteq \pc^\dprime \sqsubseteq \pc'$ and such that the resulting program configuration $\xangled{c',m',\inenv'}$ is well-formed w.r.t. $\Gamma$, $\pc^\dprime$, and $\pc'$.
\end{restatable}

To reason about what an internal attacker learns from observing a run, we define projection of program events $\alpha$ to level $\ladv$ (Definition \ref{definition:program_event_projection}) capturing the attacker observable changes to the internal state of the system. As our model distinguishes between the secrecy levels of the size of a string and its value, we extend the grammar for program events with an event capturing that a string of size $s$ was assigned to variable $x$.
\[
    \alpha \Coloneqq\;
    \ldots \mid \assignsizeevent{x}{s}
\]
As program event projection is similar to runtime event projection, we use similar notation, but annotate with a bullet to signify that these relate to internal state.
\begin{definition}[Program event projection] \label{definition:program_event_projection}
The projection of program event $\alpha$ to level $\ladv$, written $\projeventinternal{\alpha}{\ladv}$, is defined as
\begingroup
\allowdisplaybreaks
\begin{align*}
    \projeventinternal{\epsilon}{\ladv} &= \epsilon
    \\
    \projeventinternal{\scheduleevent{\ell}{n}{w}}{\ladv} &=
        \scheduleevent{\ell}{n}{w}
    \\
    \projeventinternal{\assignevent{x}{v}}{\ladv} &=
    \begin{cases}
        \assignevent{x}{v}
            &\text{if } \Gamma(x) = \ltype{\sigma}{\ell}
            \text{ s.t. } \ell \sqsubseteq \ladv\\
        \assignsizeevent{x}{s}
            &\text{if } \Gamma(x) = \ltype{\stringtype{\ell'}}{\ell}\\
            &\text{s.t. } \ell \not\sqsubseteq \ladv
            \land \ell' \sqsubseteq \ladv\\
            &\land\; \textit{size}(v) = s\\
        \epsilon
            &\text{otherwise}
    \end{cases}
    \\
    \projeventinternal{\queueevent{\ell}{v}}{\ladv} &=
    \begin{cases}
        \queueevent{\ell}{v}
            &\text{if } \ell \sqsubseteq \ladv\\
        \epsilon
            &\text{if } \ell \not\sqsubseteq \ladv
    \end{cases}
    \\
    \projeventinternal{\inputevent{\ell}{x}{v}}{\ladv} &=
    \begin{cases}
        \inputevent{\ell}{x}{v}
            &\text{if } \ell \sqsubseteq \ladv\\
        \epsilon
            &\text{if } \ell \not\sqsubseteq \ladv
    \end{cases}
\end{align*}
\endgroup
\end{definition}

We define internal trace filtering using program event projection in the straight forward way. We again filter out global events where no observable program or runtime events are emitted to prevent the attacker from observing termination.

\begin{definition}[Internal trace filtering] \label{definition:internal_trace_filtering}
The internal filtering of a trace $\tau$ at level $\ladv$, written $\filtertraceinternal{\tau}{\ladv}$, is defined as 
\begin{align*}
    \filtertraceinternal{\epsilon}{\ladv} &= \epsilon\\
    \filtertraceinternal{(\tau' \cdot (\timestamp: \alpha, \beta))}{\ladv} &= \\
    &\hspace{-5em}
    \begin{cases}
        \filtertraceinternal{\tau'}{\ladv} \cdot (\timestamp: \projeventinternal{\alpha}{\ladv}, \projevent{\beta}{\ladv})
            &\text{if } \projeventinternal{\alpha}{\ladv} \neq \epsilon\\
            &\text{or } \projevent{\beta}{\ladv} \neq \epsilon\\
        \filtertraceinternal{\tau'}{\ladv}
            &\text{otherwise}
    \end{cases}
\end{align*}
\end{definition}

Using internal trace filtering, we define internal knowledge. This definition mirrors attacker knowledge, except for using internal trace filtering, thereby giving additional power to the attacker.

\begin{definition}[Internal knowledge] \label{definition:internal_knowledge}
Given a program configuration $\lconfig$, such that $\xAngled{\lconfig,\outenv_\textit{init},\schedule_\textit{init},0}
\rightarrowdbl^*_{\tau}
\xAngled{\lconfig',\outenv',\schedule',\timestamp'}$, internal knowledge at level $\ladv$ is the set of program configurations $\lconfig_2$, that are consistent with observations at that level:
\begin{align*}
    &k^{\sbullet[1.2]}(\lconfig,\tau,\ladv) \triangleq \\
    &\quad
    \{ \lconfig_2 \mid \lconfig \ladvequiv \lconfig_2\; \land\\
    &\qquad
    \xAngled{\lconfig_2,\outenv_\textit{init},\schedule_\textit{init},0}
    \rightarrowdbl^*_{\tau_2}
    \xAngled{\lconfig'_2,\outenv'_2,\schedule'_2,\timestamp'_2}\; \land\\
    &\qquad
    (\filtertraceinternal{\tau}{\ladv}) = (\filtertraceinternal{\tau_2}{\ladv}) \}
\end{align*}
\end{definition}

Lemma \ref{lemma:external_internal_knowledge} captures that an internal attacker is indeed stronger than an external attacker by giving internal knowledge as a lower bound on attacker knowledge. This lemma allows us to relate the result we obtain for an internal attacker to the external attacker we consider in our threat model, thereby enabling us to show the type system of SELENE sound with respect to Definition \ref{definition:timing-sensitive_noninterference}.
We refer to
\ifdefined\istechnicalreport
the Appendix
\else
the accompanying technical report
\fi
for the proof.

\begin{restatable}[Internal knowledge refines external knowledge]{lemma}{externalinternalknowledge}
For any program configuration $\lconfig$, trace $\tau$, and level $\ladv$, the knowledge of an external attacker is less precise than the knowledge of an internal attacker. That is,
\label{lemma:external_internal_knowledge}
\begin{align*}
    k(\lconfig,\tau,\ladv) \supseteq k^{\sbullet[1.2]}(\lconfig,\tau,\ladv)
\end{align*}
\end{restatable}

In defining the knowledge of a network attacker in Section \ref{section:security_condition}, we defined equivalence of input environments (Definition \ref{definition:input_equivalence}) by relating only environments whose high channels receive packets at the same timestamps. However, while the network attacker observes incoming packets, they do not observe if or when the packets are consumed by the program internally. For this reason, program steps do not need to preserve equivalence by Definition \ref{definition:input_equivalence}. To relate input environments internally we define internal equivalence (Definition \ref{definition:programmatic_input_equivalence}) as equality of packet sequences on attacker observable channels.

\begin{definition}[Internal input environment equivalence up to level]  \label{definition:programmatic_input_equivalence}
Two input environments $\inenv$ and $\inenv'$ are internally equivalent up to level $\ladv$, written $\inenv \ladvequivprog \inenv'$, if
\[
\inferrule {
    \ell \sqsubseteq \ladv \implies \inenv_1(\ell) = \inenv_2(\ell) \\
} {
    \inenv_1 \ladvequivprog\inenv_2
}
\]
\end{definition}

Definition \ref{definition:programmatic_input_equivalence} is strictly weaker than Definition \ref{definition:input_equivalence}, which we state as Lemma \ref{lemma:inenvequiv_implies_inenvequivprog}.

\begin{restatable}[Network input equivalence implies internal input equivalence]{lemma}{inenvequiv_implies_inenvequivprog}
\label{lemma:inenvequiv_implies_inenvequivprog}
For any input environments $\inenv_1,\inenv_2$, equivalence by Definition \ref{definition:input_equivalence} implies equivalence by Definition \ref{definition:programmatic_input_equivalence}. That is,
\begin{align*}
    \inenv_1 \ladvequiv \inenv_2 \implies \inenv_1 \ladvequivprog \inenv_2
\end{align*}
\end{restatable}
\begin{IEEEproof}
    Immediate from the definitions.
\end{IEEEproof}

We now present our noninterference lemma for program configurations (Lemma \ref{lemma:noninterference_local}). It says that given a level $\ladv$ and a program configuration that takes a step emitting some event $\alpha$, then all configurations equivalent at $\ladv$ either take a step, emitting an equivalent event $\alpha'$, and are again equivalent at $\ladv$; or the configuration can be typed with a high $\pc$ and has no observable effects at level $\ladv$.

\begin{restatable}[Program step noninterference]{lemma}{noninterferencelocal}
\label{lemma:noninterference_local}
Given a level $\ladv$ and a command $c$ and $\Gamma,\pc,\pc'$ such that $c$ is well-formed w.r.t $\Gamma,\pc,\pc'$, if
\[
    \xangled{c,m_1,\inenv_1}
    \xrightarrow{\timestamp}_{\alpha_1}
    \xangled{c',m'_1,\inenv'_1}
\]
then for any memory $m_2$ such that $m_1 \ladvequiv m_2$ and input environment $\inenv_2$ such that $\inenv_1 \ladvequivprog \inenv_2$, we have that one of the following holds

\begin{enumerate}
    \item either $\xangled{c,m_2,\inenv_2} \xrightarrow{\timestamp}_{\alpha_2} \xangled{c',m'_2,\inenv'_2}$
    such that $\projeventinternal{\alpha_1}{\ladv} = \projeventinternal{\alpha_2}{\ladv}$ and $m'_1 \ladvequiv m'_2$ and $\inenv'_1 \ladvequivprog \inenv'_2$.
    
    \item or $\projeventinternal{\alpha_1}{\ladv} = \epsilon$ and there is $\pc^\dprime$ such that $\pc^\dprime \not\sqsubseteq \ladv$ and $\Gamma,\pc^\dprime \vdash c : \pc'$ and $m_1 \ladvequiv m'_1$ and $\inenv_1 \ladvequivprog \inenv'_1$
\end{enumerate}
\end{restatable}
We refer the interested reader to 
\ifdefined\istechnicalreport
the Appendix
\else
the accompanying technical report
\fi
for the proof and supporting lemmas.

\subsection{Global configuration}
We take the next step towards showing soundness of the type system by using Lemma \ref{lemma:noninterference_local} to show single-step noninterference of global configurations. Lemma \ref{lemma:noninterference_global} tells us that for any level $\ladv$ and any two equivalent runs, if one run takes a step producing global event $\gamma$, then the other run also takes an equivalent step, or the two runs diverge  with high $\pc$. We do not require that all events emitted after divergence are unobservable to the adversary. Instead, we allow runtime events to still be observable. This is intuitively safe as the runtime behaviour is fully determined by the program events, and no observable program events may be emitted after divergence. To simplify reasoning for divergent runs, we use a configuration with command $\texttt{stop}$ in the second condition. This configuration serves as an anchor, allowing us to reason about the observable behaviour of any two equivalent configurations by showing they behave the same as the \texttt{stop}-configuration. To satisfy this condition, the configuration must produce no observable program event or changes to memory or input environment, and produces output if and only if the \texttt{stop}-configuration produces equivalent output.

\begin{restatable}[Global step noninterference]{lemma}{noninterferenceglobal}
\label{lemma:noninterference_global}
Given a level $\ladv$ and a command $c$ and $\Gamma,\pc,\pc'$ such that $c$ is well-formed w.r.t $\Gamma,\pc,\pc'$, if
\[
    \xAngled{\xangled{c,m_1,\inenv_1},\outenv_1,\schedule,\timestamp}
    \rightarrowdbl_{(\timestamp: \alpha_1, \beta_1)}
    \xAngled{\xangled{c',m'_1,\inenv'_1},\outenv'_1,\schedule',\timestamp'}
\]
then for any memory $m_2$ such that $m_1 \ladvequiv m_2$; input environment $\inenv_2$ such that $\inenv_1 \ladvequivprog \inenv_2$; and output environment $\outenv_2$ such that $\outenv_1 \ladvequiv \outenv_2$, we have that one of the following holds

\begin{enumerate}[leftmargin=*,align=left]
    \item either
    \[
        \xAngled{\xangled{c,m_2,\inenv_2},\outenv_2,\schedule,\timestamp}
        \rightarrowdbl_{(\timestamp: \alpha_2, \beta_2)}
        \xAngled{\xangled{c',m'_2,\inenv'_2},\outenv'_2,\schedule',\timestamp'}
    \]
    such that each of the following hold
    \begin{enumerate}
        \item $\projeventinternal{\alpha_1}{\ladv} = \projeventinternal{\alpha_2}{\ladv}$
        \item $\projevent{\beta_1}{\ladv} = \projevent{\beta_2}{\ladv}$
        \item $m'_1 \ladvequiv m'_2$
        \item $\inenv'_1 \ladvequivprog \inenv'_2$
        \item $\outenv'_1 \ladvequiv \outenv'_2$
    \end{enumerate}

    \item or there is $\pc^\dprime$ such that $\pc^\dprime \not\sqsubseteq \ladv$ and $c$ is well-formed w.r.t $\Gamma,\pc^\dprime,\pc'$, and such that each of the following hold
    \begin{enumerate}
        \item $\projeventinternal{\alpha_1}{\ladv} = \epsilon$
        \item $m_1 \ladvequiv m'_1$
        \item $\inenv_1 \ladvequivprog \inenv'_1$
        \item $\schedule = \schedule'$
        \item $\beta_1 \neq \epsilon$ if and only if there are $\beta_2$ and $\outenv'_2$ such that $\projevent{\beta_1}{\ladv} = \projevent{\beta_2}{\ladv}$, $\outenv'_1 \ladvequiv \outenv'_2$, and
        \begin{align*}
            &\xAngled{\xangled{{\normalfont\texttt{stop}},m_2,\inenv_2},\outenv_2,\schedule,\timestamp}\\
        &\qquad
        \rightarrowdbl_{(\timestamp:\epsilon,\beta_2)}
        \xAngled{\xangled{{\normalfont\texttt{stop}},m_2,\inenv_2},\outenv'_2,\schedule,\timestamp'}
        \end{align*}
    \end{enumerate}
\end{enumerate}
\end{restatable}
We again refer to
\ifdefined\istechnicalreport
the Appendix
\else
the accompanying technical report
\fi
for the proof.

\subsection{Soundness of security type system}
Using the above results, we are now ready to state our soundness theorem (Theorem \ref{theorem:soundness}). It says that any well-typed program satisfies timing-sensitive, progress-sensitive noninterference (Definition \ref{definition:timing-sensitive_noninterference}).

\begin{restatable}[Soundness]{theorem}{soundness}
\label{theorem:soundness}
Given a typing environment $\Gamma$, two levels $\pc,\pc'$, and a program configuration $\lconfig$ that is well-formed w.r.t $\Gamma, \pc, \pc'$, the run $\xAngled{\lconfig,\outenv_\textit{init},\schedule_\textit{init},0} \rightarrowdbl^*_\tau \xAngled{\lconfig',\outenv',\schedule',\timestamp'}$ satisfies Definition \ref{definition:timing-sensitive_noninterference}.
\end{restatable}

The proof is by induction in the number of steps and uses Lemma \ref{lemma:noninterference_global} to infer that if the run emits an event observable by an internal attacker, then all runs of equivalent program configurations must as well, hence the internal attacker does not learn anything. Using Lemma \ref{lemma:external_internal_knowledge} we conclude that the network attacker equally does not learn anything. The full proof of Theorem \ref{theorem:soundness} is omitted here and can be found in
\ifdefined\istechnicalreport
the Appendix.
\else
the accompanying technical report.
\fi

As noted in Section \ref{section:enforcement}, Programs 5, 6, and 7 from Section \ref{section:introduction} are typeable by the typing rules and hence by Theorem \ref{theorem:soundness} we obtain a proof that they satisfy Definition \ref{definition:timing-sensitive_noninterference} and do not leak by their output behaviour.

\section{Discussion}
\label{section:discussion}

\subsection{Publicly observable traffic}
By our chosen strategy and the assumption that network activity can be eavesdropped, we arrive at a number of restrictions on how output channels can be used.
Our strategy does not permit the scheduling of new messages after the program counter has been raised. Nevertheless, Program 7 in Section \ref{section:introduction} demonstrates how dynamic scheduling is possible after receiving input, provided the input is received on a non-secret channel. This allows lower bandwidth overheads compared with constant rate padding schemes, by only scheduling traffic when needed. While it is unsurprising that it is safe to only schedule as needed while in a low context, this intuitive fact is difficult to prove correct without a principled, formal model like we present in this paper.

\subsection{Limitations and future work}
In this paper we have opted for simple and explicit packet scheduling via programmer written commands, allowing us to use the IFC system to prevent leaks from both message contents and message presence.
Our model and the primitives presented are not intended as a full solution for preventing traffic analysis attacks, but rather aim to bring attention to an as yet unsolved problem, and serve as a step towards providing practical and provably safe usage of channels susceptible to eavesdropping. A significant limitation of our model is that the progress-sensitive nature of the type system makes composition of programs difficult.
A $\pc$-declassification mechanism would alleviate this issue, but the security impact of allowing such mechanism must be fully understood. To focus our model, we have deliberately not included a $\pc$-declassification primitive in the language.

Other strategies for setting up packet schedules may be viable. In particular, we note that patterns in publicly observable input traffic may be used when deciding the shape of output traffic. However, we note that for receives with public blocking behaviour, the employed strategy should not incur significant overheads or hinder the ability to reply. As such, static pre-processing of a target program to determine a packet schedule is not viable.

\section{Related work}
\label{section:related_work}
Sabelfeld and Mantel \cite{sabelfeld-mantel-2002} also consider the problem of sending secret messages over publicly observable channels as part of a distributed program. They consider concurrent programs that communicate over low channels that are fully observable; encrypted channels where the number of messages is observable, but size and contents is not; and high channels where both message presence and contents are secret. They define a timing-sensitive security definition using strong low-bisimulation, requiring that the number of encrypted messages sent is the same between any two related runs in lockstep. The channels they consider model communication with specific endpoints at nodes -- rather than with nodes themselves -- and they do not contain dummy messages.
Consequently, the blocking behaviour of receives on encrypted channels is public, and encrypted channels in their work do not correspond with non-public channels of our work whose blocking behaviour is non-public.
The authors discuss the practical implications of different communication primitives. They argue that receives on channels that exhibit secret blocking behaviour is not secure and non-blocking receives should be used instead, while receives on channels that exhibit public blocking behaviour should use blocking receives to prevent busy waiting.

Zhang et al. \citep{DBLP:conf/pldi/ZhangAM12} propose a general language-based
mechanism for controlling timing channels based on the idea of predictive migitation~\cite{PredictiveMitigation:CCS10}. While both their approach and ours rely on the idea of scheduling observable events, they are orthogonal. A distinguishing property  of the predictive mitigation is that because of its generality the only allowed modification to program semantics is  delaying of the messages. In contrast, our approach -- where we focus on the network attacker -- allows us to use dummy messages, preventing the delays caused by mispredictions. 

As noted in Section \ref{section:discussion}, a $\pc$-declassification mechanism could be used to alleviate some of the restrictions imposed by SELENE's progress-sensitive type system. Bay and Askarov \cite{bay-askarov-2020} give a formal condition on how much attacker knowledge is allowed to change as a result of $\pc$-declassification by bounding it using the so-called progress-knowledge. We leave adapting their approach to SELENE as future work. Vassena et al. \cite{vassena-2019} propose a dynamic language-level IFC system that supports deterministic parallel thread execution. Such a system could retain a public context thread, potentially mitigating the need for explicit $\pc$-declassification.

Oblivious programming languages such as ObliVM \cite{liu-2015} and Obliv-C \cite{zahur-evans-2015} allow programmers to write protocols for secure computations, where multiple parties can perform computation collaboratively without revealing their input via produced trace, e.g. instructions, memory accesses, and values of public variables. To achieve security, such languages commonly simulate the execution of non-chosen branches in conditional statements and publicly bound and pad the number of loop iterations and the number of bits needed to represent secret values. While the goal of oblivious programming languages overlaps with ours at a high level, care is needed for adapting the techniques to our model. Generally speaking, these languages do not allow loop guards or blocking behaviour to be non-public. Our model allows the size of network messages to be kept secret by sending them as a series of (potentially dummy) packets. Consequently, the blocking behaviour of the receive primitive for non-public channels in SELENE is inherently non-public. A solution used in the oblivious approach is to tag values with a public, conservative upper bound on their size. This gives weaker confidentiality, but if acceptable appears a viable solution for the problem we discuss in this paper and we leave application of oblivious programming techniques as future work.

Previous work has examined the possibility of traffic analysis attacks revealing sensitive user information and actions across various settings.

\subsubsection*{Browsers}
Chen et al. \cite{chen-2010} and Miller et al. \cite{miller-2014} both consider the traffic patterns generated by user interactions on webpages. Miller et al. present an attack against the HTTPS deployments of industry-leading websites spanning multiple sectors. Their attack was able to identify pages within a site with high accuracy, exposing personal details including medical conditions and financial affairs. They propose a defence mechanism that pads contiguous bursts of traffic up to per-website, predefined thresholds. Their analysis shows that the proposed defence mechanism outperforms site-agnostic approaches that pad the sizes of all packets to global, nearest threshold values.
Chen et al. find that the potential for traffic analysis attacks is exacerbated by design features for dynamic, reactive websites such as AJAX GUI widgets, which often generate distinctive traffic in response to user interactions.

Cherubin et al. \cite{cherubin-2017} consider website fingerprinting defences at the application layer and introduce ALPaCA, a server side defence for use with Tor. ALPaCA works by transforming site content to conform to average site content, as analysed across multiple Tor sites. Their analysis shows that ALPaCA reduces website fingerprinting accuracy from 69.6\% to 10\%.

\subsubsection*{Phones and apps}
Conti et al. \cite{conti-2016} and Wang et al. \cite{wang-2015} both consider attacks on users of Android smartphones. Conti et al. present a machine learning assisted traffic analysis attack that infers user actions in apps with high precision and high recall, e.g., opening a profile page on Facebook or posting a message on Twitter. Wang et al. present a packet level attack on encrypted Android traffic. By collecting and analysing a small amount of wireless traffic, they are able to determine which apps smartphone users are using. Their analysis shows that apps are more susceptible to traffic analysis attacks than online services accessed over browsers, as apps tend to generate more distinct patterns of traffic.

Bahramali et al. \cite{bahramali-2020} show that also instant messaging clients are susceptible to traffic analysis attacks despite using state-of-the-art encryption. They demonstrate an attack capable of identifying members and administrators of IM channels with high accuracy, using only low-cost traffic analysis techniques. They attribute this to the fact that major IM operators do not use mechanisms for obfuscating genuine traffic, arguing their reluctance is due to the performance and usability impact of deploying such techniques.

\subsubsection*{In the home}
Zhang et al. \cite{zhang-2011} present an attack for inferring user activities by eavesdropping on WLAN traffic. They consider online activities such as web browsing, chatting, gaming, and watching videos. They use a hierarchical classification system based on machine learning algorithms and show that their system can distinguish different online applications with roughly 80\% accuracy when given 5 seconds of traffic, and roughly 90\% accuracy when given 1 minute of traffic.

Apthorpe et al. \cite{apthorpe-2018} consider home IoT devices and attacks inferring when a device is used, thereby revealing sensitive user information such as sleep patterns and when the user is home. They introduce stochastic traffic padding, which decreases attacker confidence by uniformly shaping upload and download traffic during user activities, and injecting equivalent traffic patterns at random times to hide when the device is in use.

\section{Conclusion}
\label{section:conclusion}
In this paper we consider language-based mitigation of traffic analysis attacks. We observe four traits on messages sent that may leak secret information, namely presence, recipient, size, and time. This observation informed the design of SELENE, a small imperative language for interactive programs. The type system of SELENE enforces principled, provably secure communication over channels where packets are publicly observable. The key insight of the language is a novel primitive that provides programmatic control over traffic shaping thereby allowing for reduced overheads in latency and bandwidth compared with black box techniques. We give a formal, timing-sensitive, progress-sensitive security condition based on the knowledge-based approach and prove our type system sound. We believe that our model faithfully captures online communication constraints, and that our results constitute a step towards practical, secure online communication.
We believe the security risks of traffic analysis attacks against confidentiality are significant and that work on language-based information flow for interactive programs must be mindful of the assumptions being made about the security of communications channels. We welcome and encourage future work to explore language-based techniques for providing strong security guarantees against traffic analysis attacks.

\section{Acknowledgements}
We thank Alix Trieu for his comments and feedback and thank the anonymous reviewers for their valuable suggestions for improving the presentation of this paper.
This work was funded by the Danish Council Independent Research 
for the Natural Sciences (DFF/FNU, project 6108-00363).

\bibliography{IEEEabrv,literature}

\begin{thebibliography}{10}
\providecommand{\url}[1]{#1}
\csname url@samestyle\endcsname
\providecommand{\newblock}{\relax}
\providecommand{\bibinfo}[2]{#2}
\providecommand{\BIBentrySTDinterwordspacing}{\spaceskip=0pt\relax}
\providecommand{\BIBentryALTinterwordstretchfactor}{4}
\providecommand{\BIBentryALTinterwordspacing}{\spaceskip=\fontdimen2\font plus
\BIBentryALTinterwordstretchfactor\fontdimen3\font minus
  \fontdimen4\font\relax}
\providecommand{\BIBforeignlanguage}[2]{{%
\expandafter\ifx\csname l@#1\endcsname\relax
\typeout{** WARNING: IEEEtranS.bst: No hyphenation pattern has been}%
\typeout{** loaded for the language `#1'. Using the pattern for}%
\typeout{** the default language instead.}%
\else
\language=\csname l@#1\endcsname
\fi
#2}}
\providecommand{\BIBdecl}{\relax}
\BIBdecl

\bibitem{apthorpe-2018}
\BIBentryALTinterwordspacing
N.~J. Apthorpe, D.~Y. Huang, D.~Reisman, A.~Narayanan, and N.~Feamster,
  ``Keeping the smart home private with smart(er) iot traffic shaping,''
  \emph{CoRR}, vol. abs/1812.00955, 2018. [Online]. Available:
  \url{http://arxiv.org/abs/1812.00955}
\BIBentrySTDinterwordspacing

\bibitem{askarov-chong-2012}
A.~Askarov and S.~Chong, ``Learning is change in knowledge: Knowledge-based
  security for dynamic policies,'' \emph{2012 IEEE 25th Computer Security
  Foundations Symposium}, pp. 308--322, 2012.

\bibitem{askarov-sabelfeld-2009}
A.~{Askarov} and A.~{Sabelfeld}, ``Tight enforcement of information-release
  policies for dynamic languages,'' in \emph{2009 22nd IEEE Computer Security
  Foundations Symposium}, 2009, pp. 43--59.

\bibitem{askarov-myers-2011}
\BIBentryALTinterwordspacing
A.~Askarov and A.~C. Myers, ``Attacker control and impact for confidentiality
  and integrity,'' \emph{Logical Methods in Computer Science}, vol.~7, no.~3,
  2011. [Online]. Available: \url{https://doi.org/10.2168/LMCS-7(3:17)2011}
\BIBentrySTDinterwordspacing

\bibitem{askarov2007gradual}
A.~Askarov and A.~Sabelfeld, ``Gradual release: Unifying declassification,
  encryption and key release policies,'' in \emph{2007 IEEE Symposium on
  Security and Privacy (SP'07)}.\hskip 1em plus 0.5em minus 0.4em\relax IEEE,
  2007, pp. 207--221.

\bibitem{PredictiveMitigation:CCS10}
\BIBentryALTinterwordspacing
A.~Askarov, D.~Zhang, and A.~C. Myers, ``Predictive black-box mitigation of
  timing channels,'' in \emph{Proceedings of the 17th ACM Conference on
  Computer and Communications Security}, ser. CCS '10.\hskip 1em plus 0.5em
  minus 0.4em\relax New York, NY, USA: ACM, 2010, pp. 297--307. [Online].
  Available: \url{http://doi.acm.org/10.1145/1866307.1866341}
\BIBentrySTDinterwordspacing

\bibitem{bahramali-2020}
\BIBentryALTinterwordspacing
A.~Bahramali, A.~Houmansadr, R.~Soltani, D.~Goeckel, and D.~Towsley,
  ``Practical traffic analysis attacks on secure messaging applications,''
  \emph{Proceedings 2020 Network and Distributed System Security Symposium},
  2020. [Online]. Available: \url{http://dx.doi.org/10.14722/ndss.2020.24347}
\BIBentrySTDinterwordspacing

\bibitem{bastys-2020}
I.~{Bastys}, M.~{Balliu}, T.~{Rezk}, and A.~{Sabelfeld}, ``Clockwork: Tracking
  remote timing attacks,'' in \emph{2020 IEEE 33rd Computer Security
  Foundations Symposium (CSF)}, 2020, pp. 350--365.

\bibitem{bay-askarov-2020}
J.~Bay and A.~Askarov, ``Reconciling progress-insensitive noninterference and
  declassification,'' in \emph{2020 IEEE 33rd Computer Security Foundations
  Symposium (CSF)}.\hskip 1em plus 0.5em minus 0.4em\relax IEEE, 2020, pp.
  95--106.

\bibitem{bohannon-2009}
\BIBentryALTinterwordspacing
A.~Bohannon, B.~C. Pierce, V.~Sj\"{o}berg, S.~Weirich, and S.~Zdancewic,
  ``Reactive noninterference,'' in \emph{Proceedings of the 16th ACM Conference
  on Computer and Communications Security}, ser. CCS '09.\hskip 1em plus 0.5em
  minus 0.4em\relax New York, NY, USA: Association for Computing Machinery,
  2009, p. 79–90. [Online]. Available:
  \url{https://doi.org/10.1145/1653662.1653673}
\BIBentrySTDinterwordspacing

\bibitem{chen-2010}
\BIBentryALTinterwordspacing
S.~Chen, R.~Wang, X.~Wang, and K.~Zhang, ``Side-channel leaks in web
  applications: A reality today, a challenge tomorrow,'' in \emph{Proceedings
  of the 2010 IEEE Symposium on Security and Privacy}, ser. SP '10.\hskip 1em
  plus 0.5em minus 0.4em\relax USA: IEEE Computer Society, 2010, p. 191–206.
  [Online]. Available: \url{https://doi.org/10.1109/SP.2010.20}
\BIBentrySTDinterwordspacing

\bibitem{cherubin-2017}
\BIBentryALTinterwordspacing
G.~Cherubin, J.~Hayes, and M.~Ju{\'{a}}rez, ``Website fingerprinting defenses
  at the application layer,'' \emph{PoPETs}, vol. 2017, no.~2, pp. 186--203,
  2017. [Online]. Available: \url{https://doi.org/10.1515/popets-2017-0023}
\BIBentrySTDinterwordspacing

\bibitem{clark-hunt-2009}
D.~Clark and S.~Hunt, ``Non-interference for deterministic interactive
  programs,'' in \emph{Formal Aspects in Security and Trust: 5th International
  Workshop, FAST 2008 Malaga, Spain, October 9-10, 2008 Revised Selected
  Papers}, 04 2009, pp. 50--66.

\bibitem{clarkson-scheider-2010}
\BIBentryALTinterwordspacing
M.~R. Clarkson and F.~B. Schneider, ``Hyperproperties,'' \emph{J. Comput.
  Secur.}, vol.~18, no.~6, pp. 1157--1210, Sep. 2010. [Online]. Available:
  \url{http://dl.acm.org/citation.cfm?id=1891823.1891830}
\BIBentrySTDinterwordspacing

\bibitem{conti-2016}
M.~{Conti}, L.~V. {Mancini}, R.~{Spolaor}, and N.~V. {Verde}, ``Analyzing
  android encrypted network traffic to identify user actions,'' \emph{IEEE
  Transactions on Information Forensics and Security}, vol.~11, no.~1, pp.
  114--125, 2016.

\bibitem{das-meiser-mohammadi-kate-2017}
D.~Das, S.~Meiser, E.~Mohammadi, and A.~Kate, ``Anonymity trilemma: Strong
  anonymity, low bandwidth overhead, low latency - choose two,'' \emph{{IACR}
  Cryptology ePrint Archive}, vol. 2017, p. 954, 2017.

\bibitem{dyer-2012}
\BIBentryALTinterwordspacing
K.~P. Dyer, S.~E. Coull, T.~Ristenpart, and T.~Shrimpton, ``Peek-a-boo, i still
  see you: Why efficient traffic analysis countermeasures fail,'' in
  \emph{Proceedings of the 2012 IEEE Symposium on Security and Privacy}, ser.
  SP '12.\hskip 1em plus 0.5em minus 0.4em\relax USA: IEEE Computer Society,
  2012, p. 332–346. [Online]. Available:
  \url{https://doi.org/10.1109/SP.2012.28}
\BIBentrySTDinterwordspacing

\bibitem{feigenbaum-2010}
J.~Feigenbaum, A.~Johnson, and P.~Syverson, ``Preventing active timing attacks
  in low-latency anonymous communication,'' in \emph{Privacy Enhancing
  Technologies}.\hskip 1em plus 0.5em minus 0.4em\relax Berlin, Heidelberg:
  Springer Berlin Heidelberg, 07 2010, pp. 166--183.

\bibitem{fu-2003}
X.~Fu, B.~Graham, R.~Bettati, W.~Zhao, and D.~Xuan, ``Analytical and empirical
  analysis of countermeasures to traffic analysis attacks,'' in \emph{2003
  International Conference on Parallel Processing, 2003. Proceedings.}, 11
  2003, pp. 483 -- 492.

\bibitem{goguen-meseguer-1982}
J.~A. {Goguen} and J.~{Meseguer}, ``Security policies and security models,'' in
  \emph{1982 IEEE Symposium on Security and Privacy}, 1982, pp. 11--11.

\bibitem{hedin-sabelfeld-2012}
D.~Hedin and A.~Sabelfeld, ``A perspective on information-flow control,'' in
  \emph{Software Safety and Security}, 2012.

\bibitem{juarez-2015}
\BIBentryALTinterwordspacing
M.~Ju{\'{a}}rez, M.~Imani, M.~Perry, C.~D{\'{\i}}az, and M.~Wright,
  ``{WTF-PAD:} toward an efficient website fingerprinting defense for tor,''
  \emph{CoRR}, vol. abs/1512.00524, 2015. [Online]. Available:
  \url{http://arxiv.org/abs/1512.00524}
\BIBentrySTDinterwordspacing

\bibitem{kwon-2015}
\BIBentryALTinterwordspacing
A.~Kwon, M.~AlSabah, D.~Lazar, M.~Dacier, and S.~Devadas, ``Circuit
  fingerprinting attacks: Passive deanonymization of tor hidden services,'' in
  \emph{Proceedings of the 24th USENIX Conference on Security Symposium}, ser.
  SEC'15.\hskip 1em plus 0.5em minus 0.4em\relax Berkeley, CA, USA: USENIX
  Association, 2015, pp. 287--302. [Online]. Available:
  \url{http://dl.acm.org/citation.cfm?id=2831143.2831162}
\BIBentrySTDinterwordspacing

\bibitem{liu-2015}
C.~{Liu}, X.~S. {Wang}, K.~{Nayak}, Y.~{Huang}, and E.~{Shi}, ``Oblivm: A
  programming framework for secure computation,'' in \emph{2015 IEEE Symposium
  on Security and Privacy}, 2015, pp. 359--376.

\bibitem{mantel-sabelfeld-2003}
H.~Mantel and A.~Sabelfeld, ``A unifying approach to the security of
  distributed and multi-threaded programs,'' \emph{J. Comput. Secur.}, vol.~11,
  no.~4, p. 615–676, Jul. 2003.

\bibitem{miller-2014}
B.~Miller, L.~Huang, A.~D. Joseph, and J.~D. Tygar, ``I know why you went to
  the clinic: Risks and realization of https traffic analysis,'' in
  \emph{Privacy Enhancing Technologies}, E.~De~Cristofaro and S.~J. Murdoch,
  Eds.\hskip 1em plus 0.5em minus 0.4em\relax Cham: Springer International
  Publishing, 2014, pp. 143--163.

\bibitem{oneill-clarkson-chong-2006}
\BIBentryALTinterwordspacing
K.~R. O'Neill, M.~R. Clarkson, and S.~Chong, ``Information-flow security for
  interactive programs,'' in \emph{Proceedings of the 19th IEEE Workshop on
  Computer Security Foundations}, ser. CSFW '06.\hskip 1em plus 0.5em minus
  0.4em\relax Washington, DC, USA: IEEE Computer Society, 2006, pp. 190--201.
  [Online]. Available: \url{https://doi.org/10.1109/CSFW.2006.16}
\BIBentrySTDinterwordspacing

\bibitem{overdorf-2017}
\BIBentryALTinterwordspacing
R.~Overdorf, M.~Ju{\'{a}}rez, G.~Acar, R.~Greenstadt, and C.~D{\'{\i}}az, ``How
  unique is your .onion? an analysis of the fingerprintability of tor onion
  services,'' \emph{CoRR}, vol. abs/1708.08475, 2017. [Online]. Available:
  \url{http://arxiv.org/abs/1708.08475}
\BIBentrySTDinterwordspacing

\bibitem{panchenko-2015}
A.~Panchenko, F.~Lanze, A.~Zinnen, M.~Henze, J.~Pennekamp, K.~Wehrle, and
  T.~Engel, ``Fingerprinting at internet scale,'' in \emph{Proceedings of the
  23rd Internet Society (ISOC) Network and Distributed System Security
  Symposium (NDSS 2016)}, 2015.

\bibitem{sabelfeld-mantel-2002}
A.~Sabelfeld and H.~Mantel, ``Static confidentiality enforcement for
  distributed programs,'' in \emph{Static Analysis}, M.~V. Hermenegildo and
  G.~Puebla, Eds.\hskip 1em plus 0.5em minus 0.4em\relax Berlin, Heidelberg:
  Springer Berlin Heidelberg, 2002, pp. 376--394.

\bibitem{schoepe-sabelfeld-2015}
\BIBentryALTinterwordspacing
D.~Schoepe and A.~Sabelfeld, ``Understanding and enforcing opacity,'' in
  \emph{Proceedings of the 2015 IEEE 28th Computer Security Foundations
  Symposium}, ser. CSF '15.\hskip 1em plus 0.5em minus 0.4em\relax Washington,
  DC, USA: IEEE Computer Society, 2015, pp. 539--553. [Online]. Available:
  \url{https://doi.org/10.1109/CSF.2015.41}
\BIBentrySTDinterwordspacing

\bibitem{siby-2018}
S.~Siby, M.~Ju{\'a}rez, N.~Vallina-Rodriguez, and C.~Troncoso, ``Dns privacy
  not so private: the traffic analysis perspective,'' 2018.

\bibitem{vassena-2019}
M.~Vassena, G.~Soeller, P.~Amidon, M.~Chan, J.~Renner, and D.~Stefan,
  ``Foundations for parallel information flow control runtime systems,'' in
  \emph{Principles of Security and Trust}, F.~Nielson and D.~Sands, Eds.\hskip
  1em plus 0.5em minus 0.4em\relax Cham: Springer International Publishing,
  2019, pp. 1--28.

\bibitem{wang-2015}
Q.~{Wang}, A.~{Yahyavi}, B.~{Kemme}, and W.~{He}, ``I know what you did on your
  smartphone: Inferring app usage over encrypted data traffic,'' in \emph{2015
  IEEE Conference on Communications and Network Security (CNS)}, 2015, pp.
  433--441.

\bibitem{zahur-evans-2015}
S.~Zahur and D.~Evans, ``Obliv-c: A language for extensible data-oblivious
  computation,'' Cryptology ePrint Archive, Report 2015/1153, 2015,
  \url{https://eprint.iacr.org/2015/1153}.

\bibitem{DBLP:conf/pldi/ZhangAM12}
\BIBentryALTinterwordspacing
D.~Zhang, A.~Askarov, and A.~C. Myers, ``Language-based control and mitigation
  of timing channels,'' in \emph{{ACM} {SIGPLAN} Conference on Programming
  Language Design and Implementation, {PLDI} '12, Beijing, China - June 11 -
  16, 2012}, 2012, pp. 99--110. [Online]. Available:
  \url{https://doi.org/10.1145/2254064.2254078}
\BIBentrySTDinterwordspacing

\bibitem{zhang-2011}
\BIBentryALTinterwordspacing
F.~Zhang, W.~He, X.~Liu, and P.~G. Bridges, ``Inferring users' online
  activities through traffic analysis,'' in \emph{Proceedings of the Fourth ACM
  Conference on Wireless Network Security}, ser. WiSec '11.\hskip 1em plus
  0.5em minus 0.4em\relax New York, NY, USA: Association for Computing
  Machinery, 2011, p. 59–70. [Online]. Available:
  \url{https://doi.org/10.1145/1998412.1998425}
\BIBentrySTDinterwordspacing

\end{thebibliography}

\ifdefined\istechnicalreport
\clearpage

\onecolumn
\appendix
\subsection{Local configuration}
\label{appendix:local}
\begin{lemma}[Progress-sensitive typing] \label{lemma:progress_sensitive_typing}
Let $\Gamma$ be a typing environment, $\pc,\pc'$ two levels and $c$ a command. If $\;\Gamma,\pc \vdash c : \pc'$ then $\pc \sqsubseteq \pc'$.
\end{lemma}
\begin{IEEEproof}
    By induction on $c$ and monotinicity of $\sqsubseteq$.
\end{IEEEproof}

\preservationwellformedness*
\begin{IEEEproof}
The proof proceeds by induction on the structure of $c$.

\begin{description}
    \item \case{\texttt{skip}} By $\textsc{Skip}$ we observe that $c'$ is $\texttt{stop}$ and $m = m'$, hence we are done.
    
    \item \case{$c_1;c_2$}
    We case analyse on whether $c' = c_2$. If $c' = c_2$, we observe that $c'$ was produced using $\textsc{Seq-2}$ was taken and are done by the induction hypothesis, $\textsc{T-Seq}$, Lemma \ref{lemma:progress_sensitive_typing}. If $c' \neq c_2$, we observe that $c'$ was produced using $\textsc{Seq-1}$ and are done by the induction hypothesis and $\textsc{T-Seq}$.
    
    \item \case{\texttt{x = e}}
    By $\textsc{Assign}$ we have that $c'$ is $\texttt{stop}$ and $m' = m[x \mapsto v]$ where $\eval{e}{m}{v}$. Observing that \textsc{T-Assign} is the only applicable typing rule we have that $x \in \textsf{dom}(\Gamma)$ and $t_e <: t_x$ by which we conclude that $m$ is well-formed w.r.t. $\Gamma$ and we are done.
    
    \item \case{$x = \texttt{sizeof}(e)$}
    By \textsc{SizeOf} we observe $c'$ is $\texttt{stop}$ and $m' = m[x \mapsto n]$ where $\eval{e}{m}{v}$ and $n = \left\lceil \frac{\textit{size}(v)}{\eta} \right\rceil$ and hence $n \in \textit{Int}$. By $\textsc{T-SizeOf}$ we have $\Gamma \vdash x : \ltype{\text{Int}}{\ell_x}$, and hence $m$ is well-formed w.r.t. $\Gamma$ and we are done.
    
    \item \case{$\texttt{await}(r)$}
    By $\textsc{Await}$ we observe that $c'$ is $\texttt{stop}$ and $m' = m$, hence we are done.
    
    \item \case{$\texttt{sleep}(e)$}
    By $\textsc{Sleep}$ we observe that $c'$ is $\texttt{await}(r)$ for some $r$ and $m' = m$. As command $\texttt{await}$ is trivially typed, we are done.
    
    \item \case{$\texttt{if } e \texttt{ then } c_1 \texttt{ else } c_2$}
    We observe that $m' = m$ as neither $\textsc{If-T}$ nor $\textsc{If-E}$ alters the memory.
    We proceed by case analysis of $\eval{e}{m}{0}$, in both cases we are done by applying $\textsc{T-If}$.
    
    \item \case{$\texttt{while } e \texttt{ do } c$}
    We observe that $\textsc{While}$ transitions to a conditional branching with $m' = m$. For the first branch, we observe that by $\textsc{T-While}$ we have $\Gamma \vdash e: \ltype{\text{Int}}{\ell}$ and $\Gamma,\pc \sqcup \ell \vdash c : \pc'$ hence we are done by $\textsc{T-If}$ using the induction hypothesis. For the other branch we are done by $\textsc{T-If}$ and $\textsc{T-Skip}$ using the induction hypothesis.
    
    \item \case{$x = \texttt{in}(\ell)$}
    By $\textsc{In}$ we have that $c'$ is $\texttt{stop}$ and $m(x) \in A$. We also have $m' = m[x \mapsto v]$ for some $v \in A$ by definition of the $\textsf{choose}$ function. Hence $m'$ is well-type w.r.t. $\Gamma$ and we are done.
    
    \item \case{$\texttt{schedule}(\ell,e_1,e_2)$}
    By $\textsc{Schedule}$ we observe that $c'$ is $\texttt{stop}$ and $m' = m$, hence we are done.
    
    \item \case{$\texttt{queue}(\ell,e)$}
    By $\textsc{Queue}$ we observe that $c'$ is $\texttt{stop}$ and $m' = m$, hence we are done.
\end{description}
\end{IEEEproof}


\begin{restatable}[High program steps have only high effects]{lemma}{highstepshigheffects}
\label{lemma:high_steps_high_effects}
Given an attacker level $\ladv$, a typing environment $\Gamma$, two levels $\pc, \pc'$ such that $\pc \not\sqsubseteq \ladv$, and a program configuration $\xangled{c,m,\inenv}$ such that $\Gamma, \pc \vdash c : \pc'$ and a timestamp $\timestamp$ such that
\[
    \xangled{c,m,\inenv}
    \xrightarrow{\timestamp}_{\alpha}
    \xangled{c',m',\inenv'}
\]
then $\projeventinternal{\alpha}{\ladv} = \epsilon$ and $m \ladvequiv m'$ and $\inenv \ladvequivprog \inenv'$.
\end{restatable}
\begin{IEEEproof}
We proceed by induction on command $c$.

\begin{description}
    \item \case{\texttt{skip}}
    Trivially done by $\textsc{Skip}$, observing that $m = m'$, $\inenv = \inenv'$ and $\alpha = \epsilon$.
    
    \item \case{$c_1;c_2$}
    We case on whether $c' = c_2$. If false, we observe that the only matching rule in the operational semantics is $\textsc{Seq-1}$ and are done by applying the induction hypothesis. If true, we observe that the only matching rule in the operational semantics is $\textsc{Seq-2}$ and we are done by applying the induction hypothesis.
    
    \item \case{\texttt{x = e}}
    We observe by $\textsc{Assign}$ that $\inenv = \inenv'$ and have that $\alpha = \assignevent{x}{v}$ and $m' = m[x \mapsto v]$. We must show $m \ladvequiv m'$ and $\projeventinternal{\assignevent{x}{v}}{\ladv} = \epsilon$.
    
    By \textsc{T-Assign} we have (a) $\Gamma \vdash = \ltype{\sigma_e}{\ell_e}$, (b) $\Gamma(x) = \ltype{\sigma_x}{\ell_x}$, (c) $\sigma_e \nearrow \pc <: \sigma_t$ and (d) $\ell_e \sqcup \pc \sqsubseteq \ell_x$, hence in particular $\ell_x \not\sqsubseteq \ladv$.  We case on $\sigma_x$ to pick the corresponding projection. If $\sigma_x = \inttype$ we have $\projeventinternal{\alpha}{\ladv} = \epsilon$ and are done by observing that condition both conditions for memory equivalence is satisfied hence $m \ladvequiv m'$. If $\sigma_x = \stringtype{\ell'}$ we have by definition of $\nearrow$ and $<:$ that $\ell' \not\sqsubseteq \ladv$ hence $\projeventinternal{\alpha}{\ladv} = \epsilon$ and we are done by observing $m \ladvequiv m'$ by definition of memory equivalence.
    
    \item \case{$x = \texttt{sizeof}(e)$}
     We observe by $\textsc{SizeOf}$ that $\inenv = \inenv'$ and $\alpha = \assignevent{x}{n}$ where $n = \left\lceil \frac{\textit{size}(v)}{\eta} \right\rceil$ for some $v$ such that $\xangled{e, m} \Downarrow v$ and $m' = m[x \mapsto v]$. We must show $m \ladvequiv m'$ and $\projeventinternal{\assignevent{x}{n}}{\ladv} = \epsilon$.
     
     By \textsc{T-SizeOf} we have (a) $\Gamma \vdash x: \ltype{\inttype}{\ell_x}$, (b) $\Gamma \vdash e : \ltype{\sigma_e}{\ell_e}$, (c) $\pc \sqsubseteq \ell_x$, and (d) $\sigma_e = \stringtype{\ell'} \implies \ell' \sqsubseteq \ell_x$. By (a) and the definition of memory equivalence we have $m \ladvequiv m'$. By (c) and assumption $\pc \not\sqsubseteq \ladv$ we have (e) $\ell_x \not\sqsubseteq \ladv$ hence by definition of internal event projection we have $\projeventinternal{\assignevent{x}{n}}{\ladv} = \epsilon$ and we are done.
    
    \item \case{$\texttt{await}(r)$}
    Trivially done by $\textsc{Await}$, observing that $m = m'$, $\inenv = \inenv'$ and $\alpha = \epsilon$.
    
    \item \case{$\texttt{sleep}(e)$}
    Trivially done by $\textsc{Sleep}$, observing that $m = m'$, $\inenv = \inenv'$ and $\alpha = \epsilon$.
    
    \item \case{$\texttt{if } e \texttt{ then } c_1 \texttt{ else } c_2$}
    We case on $\eval{e}{v}{0}$. If true then by $\textsc{If-T}$, we observe that $m = m'$, $\inenv = \inenv'$, and $\alpha = \epsilon$ and we are done. Analogously by $\textsc{If-E}$ if false.
    
    \item \case{$\texttt{while } e \texttt{ do } c$}
    Trivially done by $\textsc{While}$, observing that $m = m'$, $\inenv = \inenv'$, and $\alpha = \epsilon$.
    
    \item \case{$x = \texttt{in}(\ell)$}
    We observe by $\textsc{In}$ that $\inenv(\ell)=\vec{p}$, $\textsf{choose}(\vec{p},A,\timestamp,[])=(v,\vec{q})$, $\alpha = \inputevent{\ell}{x}{v}$, $m' = m[x \mapsto v]$, and $\inenv' = \inenv[\ell \mapsto \vec{q}]$. We must show $\projeventinternal{\inputevent{\ell}{x}{v}}{\ladv} = \epsilon$, $m \ladvequiv m'$, and $\inenv \ladvequivprog \inenv'$.
    
    By $\textsc{T-In}$ we have (a) $\Gamma \vdash x : \ltype{\sigma_x}{\ell_x}$, (b) $\pc \sqsubseteq \ell$, (c) $\sigma_x \nearrow \ell <: \sigma_x$, and (d) $\ell \sqsubseteq \ell_x$. By (b) and $\pc \not\sqsubseteq \ladv$ we have (e) $\ell \not\sqsubseteq \ladv$ and hence by definition of internal event projection have $\projeventinternal{\inputevent{\ell}{x}{v}}{\ladv} = \epsilon$ and by definition of the \textsf{choose} and input environment equivalence have $\inenv \ladvequivprog \inenv'$. To show memory equivalence, we observe that by (d) and we have $\ell_x \not\sqsubseteq \ladv$ and the first condition of memory equivalence is satisfied. To see that the second condition is satisfied we case on $\sigma_x = \stringtype{\ell'}$. If false, we are done. If true, we have by (c) and the definition of $\nearrow$ and $<:$ that $\ell \sqsubseteq \ell'$ and hence $\ell' \not\sqsubseteq \ladv$ hence the second condition is satisfied and we have $m \ladvequiv m'$ and we are done.
    
    \item \case{$\texttt{schedule}(\ell,e_1,e_2)$}
    By $\textsc{T-Schedule}$ we have $\pc = \bot$, leading to contradiction $\bot \not\sqsubseteq \ladv$, hence this case is impossible.
    
    \item \case{$\texttt{queue}(\ell,e)$}
    We observe by $\textsc{Queue}$ that $m = m'$, $\inenv = \inenv'$, and we have that $\alpha = \queueevent{\ell}{v}$ such that $\eval{m}{e}{v}$.
    We must show $\outenv \ladvequiv \outenv'$ and $\projeventinternal{\queueevent{\ell}{v}}{\ladv} = \epsilon$.
    
    By $\textsc{T-Queue}$ we have (b) $\Gamma \vdash e : \ltype{\sigma_e}{\ell_e}$ and (c) $\ell_e \sqcup \ell \sqsubseteq \ell$. By (c) and assumption $\pc \not\sqsubseteq \ladv$ we have (d) $\ell \not\sqsubseteq \ladv$, hence by definition of internal event projection we have $\projeventinternal{\queueevent{\ell}{v}}{\ladv} = \epsilon$.
\end{description}
\end{IEEEproof}


\begin{lemma}[Noninterference for expressions] \label{lemma:noninterferce_expressions}
Given an attacker level $\ladv$, a typing environment $\Gamma$ and two memories $m_1$ and $m_2$ such that $m_1 \ladvequiv m_2$, and an expression $e$ such that $\Gamma \vdash e : \ltype{\sigma}{\ell}$, such that $\eval{e}{m_1}{v_1}$ and $\eval{e}{m_2}{v_2}$, then we have that
\begin{itemize}
    \item $\ell \sqsubseteq \ladv \implies v_1 = v_2$
    \item $\sigma = \stringtype{\ell'} \land \ell' \sqsubseteq \ladv \implies \textit{size}(v_1) = \textit{size}(v_2)$.
\end{itemize}
\end{lemma}
\begin{IEEEproof}
    By induction on typing derivation $\Gamma \vdash e : \ltype{\sigma}{\ell}$ using the definition of $m_1 \ladvequiv m_2$.
\end{IEEEproof}

\noninterferencelocal*
\begin{IEEEproof}
    The proof is by induction on the structure of command $c$.
    
    \begin{description}
    \item \case{$\texttt{skip}$} 
        Trivially possible for both runs in with $c' = {\normalfont\texttt{stop}}$ and $\alpha_1 = \alpha_2 = \epsilon$. We are done by observing that no updates are made to memory or input environment.
        
        \item \case{$c_1;c_2$}
        There must be $c'_1$ such that 
        \[
            \xangled{c_1,m_1,\inenv_1}
            \xrightarrow{\timestamp}_{\alpha_1}
            \xangled{c'_1,m'_1,\inenv'_1}
        \]
        and hence
        \[
            \xangled{c_1;c_2,m_1,\inenv_1}
            \xrightarrow{\timestamp}_{\alpha_1}
            \xangled{d,m'_1,\inenv'_1}
        \]
        where
        \[
            d =
            \begin{cases}
                c_2
                    &\text{if } c'_1 = \texttt{stop}\\
                c'_1;c_2
                    &\text{otherwise}
            \end{cases}
        \]
        By \textsc{T-Seq} we have $\Gamma, \pc \vdash c_1 : \pc^\dprime$ and $\Gamma, \pc^\dprime \vdash c_2 : \pc'$. By induction hypothesis we have two cases for the second run. Either we have
        \[
            \xangled{c_1;c_2,m_2,\inenv_2}
            \xrightarrow{\timestamp}_{\alpha_2}
            \xangled{d,m'_2,\inenv'_2}
        \]
        such that $m'_1 \ladvequiv m'_2$ and $\inenv'_1 \ladvequivprog \inenv'_2$ and $\projeventinternal{\alpha_1}{\ladv} = \projeventinternal{\alpha_2}{\ladv}$ and we are done by the first condition. Or we have that $m_1 \ladvequiv m'_1$ and $\inenv_1 \ladvequivprog \inenv'_1$ and $\projeventinternal{\alpha_1}{\ladv} = \epsilon$, and existence of $\pc^\trprime$ such that $\pc^\trprime \not\sqsubseteq \ladv$ and $\Gamma;\pc^\trprime \vdash c_1 : \pc^\dprime$. By this we have that $\Gamma;\pc^\trprime \vdash c_1;c_2 : \pc'$ and we are done by the second condition.
        
        \item \case{$x=e$}
        Trivially possible for both runs in one with $c' = {\normalfont\texttt{stop}}$. We observe that no updates are made to input environment.
            
        By $\textsc{T-Assign}$ we have $\Gamma \vdash x : \ltype{\sigma_x}{\ell_x}$ and $\Gamma \vdash e : \ltype{\sigma_e}{\ell_e}$ such that $\sigma_e \nearrow \pc <: \sigma_x$ and $\ell_e \sqcup \pc \sqsubseteq \ell_x$. As by $\textsc{Assign}$ we have $\alpha_1 = \assignevent{x}{v_1}$ s.t. $\eval{e}{m_1}{v_1}$ we consider the three cases for the internal event projection.
        
        If $\projeventinternal{\assignevent{x}{v_1}}{\ladv} = \assignevent{x}{v_1}$ then by definition of internal event projection we have $\ell_x \sqsubseteq \ladv$ and hence by Lemma \ref{lemma:noninterferce_expressions} have $\eval{e}{m_2}{v_2}$ s.t. $v_1 = v_2$ and hence $\alpha_2 = \alpha_1$. As $m'_1 = m_1[x \mapsto v_1]$ and $m'_2 = m_2[x \mapsto v_2]$ we are done by definition of memory equivalence.
        
        If $\projeventinternal{\assignevent{x}{v_1}}{\ladv} = \assignsizeevent{x}{s_1}$ where $s_1 = \textit{size}(v_1)$ then by definition of internal event projection we have $\sigma_e = \stringtype{\ell'}$ such that  $\ell' \sqsubseteq \ladv$ and hence by Lemma \ref{lemma:noninterferce_expressions} have $\eval{e}{m_2}{v_2}$ s.t. $\textit{size}(v_1) = \textit{size}(v_2)$ and hence $\projeventinternal{\alpha_2}{\ladv} = \projeventinternal{\alpha_1}{\ladv}$. As $m'_1 = m_1[x \mapsto v_1]$ and $m'_2 = m_2[x \mapsto v_2]$ we are done by definition of memory equivalence.
        
        If $\projeventinternal{\assignevent{x}{v_1}}{\ladv} = \epsilon$ then by definition of internal event projection we have $\ell_x \not\sqsubseteq \ladv$ and hence $\projeventinternal{\alpha_2}{\ladv} = \projeventinternal{\alpha_1}{\ladv}$. We are done by definition of memory equivalence.
        
        \item \case{$x=\texttt{sizeof}(e)$}
        Trivially possible for both runs in one with $c' = {\normalfont\texttt{stop}}$. We observe that no updates are made to input environment.
            
        By $\textsc{T-SizeOf}$ we have $\Gamma \vdash x : \ltype{\inttype}{\ell_x}$ and $\Gamma \vdash e : \ltype{\sigma_e}{\ell_e}$ such that $\ell_e \sqcup \pc \sqsubseteq \ell_x$. As by $\textsc{SizeOf}$ we have $\alpha_1 = \assignevent{x}{n_1}$ s.t. $\eval{e}{m_1}{v_1}$ and $n_1 = \left\lceil \frac{\textit{size}(v_1)}{\eta} \right\rceil$ we consider the two cases for the internal event projection.
        
        If $\projeventinternal{\assignevent{x}{n_1}}{\ladv} = \assignevent{x}{n_1}$ then by definition of the internal event projection we have $\ell_x \sqsubseteq \ladv$. By \textsc{T-SizeOf} we have that if $\sigma_e = \stringtype{\ell'}$ then $\ell' \sqsubseteq \ell_x$ and by transitivity $\ell' \sqsubseteq \ladv$. By Lemma \ref{lemma:noninterferce_expressions} we have $\eval{e}{m_2}{v_2}$ such that $\textit{size}(v_1) = \textit{size}(v_2)$, and hence $\projeventinternal{\alpha_2}{\ladv} = \projeventinternal{\alpha_1}{\ladv}$. We are done by the definition of memory equivalence.
        
        If $\projeventinternal{\assignevent{x}{n_1}}{\ladv} = \epsilon$ then by definition of internal event projection we have $\ell_x \not\sqsubseteq \ladv$ and hence $\projeventinternal{\alpha_2}{\ladv} = \projeventinternal
        {\alpha_1}{\ladv}$. We are done by definition of memory equivalence.
        
        \item \case{$\texttt{await}(r)$}
        It must be that $r \leq \timestamp$ hence this transition is possible in both runs with $c' = {\normalfont{\texttt{stop}}}$. As no event is emitted, we are done.
        
        \item \case{$\texttt{sleep}(e)$}
        If $\pc' \sqsubseteq \ladv$ then by \textsc{Sleep} we have that if $\eval{e}{m_1}{w}$ then
        \[
            \xangled{\texttt{sleep}(e),m_1,\inenv_1}
            \xrightarrow{\timestamp}_{\epsilon}
            \xangled{\texttt{await}(\timestamp+w),m_1,\inenv_1}
        \]
        By \textsc{T-Sleep} we have $\Gamma \vdash e : \ltype{\inttype}{\ell}$ and $\pc' = \pc \sqcup \ell$ and by assumption $\pc' \sqsubseteq \ladv$ we therefore have $\ell \sqsubseteq \ladv$. By Lemma \ref{lemma:noninterferce_expressions} this gives us $\eval{e}{m_2}{w}$ and hence
        \[
            \xangled{\texttt{sleep}(e),m_2,\inenv_2}
            \xrightarrow{\timestamp}_{\epsilon}
            \xangled{\texttt{await}(\timestamp+w),m_2,\inenv_2}
        \]
        
        If $\pc' \not\sqsubseteq \ladv$, we observe that by \textsc{T-Sleep} we have $\Gamma,\pc' \vdash \texttt{sleep}(e) : \pc'$ and we are done observing that by \textsc{Sleep}, $\alpha_1 = \epsilon$, and memory and input are unchanged.
        
        \item \case{$\texttt{if } e \texttt{ then } c_1 \texttt{ else } c_2$}
        By \textsc{T-If} we have $\Gamma \vdash e : \ltype{\inttype}{\ell}$ and $\Gamma;\pc \sqcup \ell \vdash c_1 : \pc_1$ and $\Gamma;\pc \sqcup \ell \vdash c_2 : \pc_2$ such that $\pc' = \pc_1 \sqcup \pc_2$. If $\ell \sqcup \ladv$, then by Lemma \ref{lemma:noninterferce_expressions} we have some $v$ such that $\eval{e}{m_1}{v}$ and $\eval{e}{m_2}{v}$, hence both runs transition to the same command and we are done by observing $\alpha_1 = \alpha_2 = \epsilon$ and memory and input environments are unchanged. If $\ell \not\sqsubseteq \ladv$ we observe again by \textsc{T-If} that $\Gamma, \pc \sqcup \ell \vdash \texttt{if } e \texttt{ then } c_1 \texttt{ else } c_2 : \pc'$ and we are done.
        
        \item \case{$\texttt{while } e \texttt{ do } c$}
        Trivial.
        
        \item \case{$x=\texttt{in}(\ell)$}
        By \textsc{T-In} we observe that $\pc' = \ell$. If $\ell \sqsubseteq \ladv$, then by definition of input environment equivalence and function $\textsf{choose}$ we have that both runs obtain same input value at the same time, therefore producing same events and making same updates to the input environment. We are done by definitions of memory equivalence and input environment equivalence.
        
        If $\ell \not\sqsubseteq \ladv$, we observe that by \textsc{T-In} we have $\Gamma,\ell \vdash x=\texttt{in}(\ell) : \ell$. By \textsc{In} we have some $v, \vec{q}$ such that $\alpha_1 = \inputevent{\ell}{x}{v}$, $m'_1 = m_1[x \mapsto v]$ and $\inenv'_1 = \inenv_1[\ell \mapsto \vec{q}]$ and by definition of the $\textsf{choose}$ function have $\inenv_1 \ladvequivprog \inenv'_1$. By definition internal event projection we have $\projeventinternal{\alpha_1}{\ladv} = \epsilon$ and we are done by definition of memory, satisfying the second condition.
        
        \item \case{$\texttt{schedule}(\ell,e_1,e_2)$}
        Trivially possible in both runs updating only schedules.
        By $\textsc{T-Schedule}$ we have $\Gamma \vdash e_1 : \ltype{\inttype}{\bot}$ and $\Gamma \vdash e_2 : \ltype{\inttype}{\bot}$ hence by Lemma \ref{lemma:noninterferce_expressions} we have $a,w$ such that $\eval{e_1}{m_1}{a}$ and $\eval{e_1}{m_2}{a}$, and $\eval{e_2}{m_1}{w}$ and $\eval{e_2}{m_2}{w}$. As $\alpha_1 = \scheduleevent{\ell}{a}{w} = \alpha_2$ we are done.
        
        \item \case{$\texttt{queue}(\ell,e)$}
        Trivially possible in both runs updating only output environments. By \textsc{Queue} we have $\eval{e}{m_1}{v_1}$ and $\alpha_1 = \queueevent{\ell}{v_1}$. We consider the two cases of the internal event projection.
        
        If $\projeventinternal{\queueevent{\ell}{v_1}}{\ladv} = \queueevent{\ell}{v_1}$ then by definition of the internal event projection we have $\ell \sqsubseteq \ladv$. By \textsc{T-Queue} we have $\Gamma \vdash e : \ltype{\sigma_e}{\ell_e}$ and $\ell_e \sqcup \pc \sqsubseteq \ell$ hence in particular $\ell_e \sqsubseteq \ladv$ and therefore by Lemma \ref{lemma:noninterferce_expressions} we have $\eval{e}{m_2}{v_2}$ such that $v_1 = v_2$ by which we have $\alpha_2 = \alpha_1$ and we are done.
        
        If $\projeventinternal{\queueevent{\ell}{v_1}}{\ladv} = \epsilon$ then by definition of internal event projection we have $\ell \not\sqsubseteq \ladv$, and therefore also $\projeventinternal{\alpha_2}{\ladv} = \epsilon$ and we are done.
    \end{description}
\end{IEEEproof}

\subsection{Global configuration} \label{appendix:global}

\begin{lemma}[High update preserves equivalence] \label{lemma:high_update_preserves_equivalence}
    For any level $\ell,\ladv$, program event $\alpha$, output environments $\outenv,\outenv'$, and schedule $\schedule$,
    if $\projeventinternal{\alpha}{\ladv} = \epsilon$ and ${\normalfont\textsf{upd}}(\outenv,\schedule,\alpha)=(\outenv',\schedule')$ then $\outenv \ladvequiv \outenv'$ and $\schedule' = \schedule$.
\end{lemma}
\begin{IEEEproof}
    By definition of $\textsf{upd}$, internal event projection and output environment equivalence.
\end{IEEEproof}

\begin{lemma}[Sending preserves output environment equivalence] \label{lemma:sending_preserves_equivalence}
    For any levels $\ell,\ladv$, schedule $\schedule$, timestamp $\timestamp$, and output environments $\outenv_1,\outenv_2$ such that $\outenv_1 \ladvequiv \outenv_2$, if $\schedule(\timestamp) = \ell$, ${\normalfont\textsf{send}}(\outenv_1,\ell) = (\beta_1,O'_1)$ and ${\normalfont\textsf{send}}(\outenv_2,\ell) = (\beta_2,O'_2)$
    then $\outenv'_1 \ladvequiv \outenv'_2$ and $\projevent{\beta_1}{\ladv} = \projevent{\beta_2}{\ladv}$.
\end{lemma}
\begin{IEEEproof}
    If $\ell \sqsubseteq \ladv$ then by definition of output environment equivalence we have $\outenv_1(\ell) = \outenv_2(\ell)$ hence $\textsf{send}(\outenv_1,\ell)=\textsf{send}(\outenv_2,\ell)$ which concludes the case.
    
    If (a) $\ell \not\sqsubseteq \ladv$, then there is $p_1,\vec{q_1}$ such that $\textsf{send}(\outenv_1,\ell)=(\outputevent{\ell}{p_1},\vec{q_1})$ and $p_2,\vec{q_2}$ such that $\textsf{send}(\outenv_2,\ell)=(\outputevent{\ell}{p_2},\vec{q_2})$. By (a) and by definition of event projection, we have $\projevent{\outputevent{\ell}{p_1}}{\ladv}=\outputevent{\ell}{-}=\projevent{\outputevent{\ell}{p_2}}{\ladv}$ and by (a) and definition of output environment equivalence we have $\outenv_1[\ell \mapsto \vec{q_1}] \ladvequiv \outenv_2[\ell \mapsto \vec{q_2}]$ which concludes the case.
\end{IEEEproof}

\noninterferenceglobal*
\begin{IEEEproof}
    We case on the global step for the first run.
    \begin{description}
        \item \case{\textsc{G-Step}}
            By Lemma \ref{lemma:noninterference_local} we have
            \begin{enumerate}
                \item[(1)] either $\xangled{c,m_2,\inenv_2} \xrightarrow{\timestamp}_{\alpha_2} \xangled{c',m'_2,\inenv'_2}$ such that $\projeventinternal{\alpha_1}{\ladv} = \projeventinternal{\alpha_2}{\ladv}$ and $m'_1 \ladvequiv m'_2$ and $\inenv'_1 \ladvequivprog \inenv'_2$
                \item[(2)] or $\projeventinternal{\alpha_1}{\ladv} = \epsilon$ and there is $\pc^\dprime$ such that $\pc^\dprime \not\sqsubseteq \ladv$ and $\Gamma;\pc^\dprime \vdash c : \pc'$ and $m_1 \ladvequiv m'_1$ and $\inenv_1 \ladvequivprog \inenv'_1$
            \end{enumerate}
            If (1), we inspect $\projeventinternal{\alpha_1}{\ladv}$. If $\projeventinternal{\alpha_1}{\ladv}=\epsilon$ then we apply Lemma \ref{lemma:high_update_preserves_equivalence} in both runs to get $\outenv'_1 \ladvequiv \outenv_1 \ladvequiv \outenv_2 \ladvequiv \outenv'_2$. We are done by Lemma \ref{lemma:sending_preserves_equivalence} satisfying condition 1. If $\projeventinternal{\alpha_1}{\ladv}\neq\epsilon$ we are done by definition of $\textsf{upd}$, observing that the same update is made in each run, and by Lemma \ref{lemma:sending_preserves_equivalence}.
            
            If (2), we inspect $\timestamp \in \textsf{dom}(\schedule)$.
            
            If $\timestamp \not\in \textsf{dom}(\schedule)$ then as $\projeventinternal{\alpha_1}{\ladv}=\epsilon$ we get by Lemma \ref{lemma:high_update_preserves_equivalence} that $\outenv_1 \ladvequiv \outenv'_1$. Further, we have by definition of internal event projection and definition of \textsf{upd} that $\schedule' = \schedule$, hence $\timestamp \not\in \textsf{dom}(\schedule')$ and by definition of \textsf{send} we have $\beta_1 = \epsilon$ and we are done satisfying condition 2.
            
            If $\schedule(\timestamp)=\ell$ we show that we satisfy condition 2 and consider each direction
            \begin{description}
                \item[$(\rightarrow)$]
                    As the first run steps by \textsc{G-Step} we have by Lemma \ref{lemma:high_update_preserves_equivalence} that $\textsf{upd}(\outenv_1,\schedule,\alpha)=(\outenv^\dprime_1,\schedule)$ such that $\outenv_1 \ladvequiv \outenv^\dprime_1$, and by Lemma \ref{lemma:sending_preserves_equivalence} have
                    $\textsf{send}(\outenv^\dprime_1,\ell)=(\beta_1,\outenv'_1)$ and
                    $\textsf{send}(\outenv_2,\ell)=(\beta_2,\outenv'_2)$ such that $\outenv'_1 \ladvequiv \outenv'_2$ and $\projevent{\beta_1}{\ladv}=\projevent{\beta_2}{\ladv}$
                    and hence
                    \[
                        \xAngled{\xangled{{\normalfont\texttt{stop}},m_2,\inenv_2},\outenv_2,\schedule,\timestamp}
                        \rightarrowdbl_{(\timestamp:\epsilon,\beta_2)}
                        \xAngled{\xangled{{\normalfont\texttt{stop}},m_2,\inenv_2},\outenv'_2,\schedule,\timestamp'}
                    \]
                    and we are done.
                \item[$(\leftarrow)$] 
                    As $\beta_2 \neq \epsilon$, the anchor-configuration steps \textsc{G-Stop} and we apply Lemma \ref{lemma:high_update_preserves_equivalence} in the first run and Lemma \ref{lemma:sending_preserves_equivalence} in both runs to get $\projevent{\beta_1}{\ladv} = \projevent{\beta_2}{\ladv}$ and $\outenv'_1 \ladvequiv \outenv'_2$ and we are done.
            \end{description}
        
        \item \case{\textsc{G-Block}}
            We have $\alpha_1 = \epsilon$ and $m_1 = m'_1$ and $\inenv_1 = \inenv'_1$ and $\outenv_1 = \outenv'_1$ and $\schedule = \schedule'$. We consider the possible steps of the second run.
            
            \begin{description}
                \item \case{\textsc{G-Step}}
                We will show that condition 2 is satisfied. To this end, we first apply the induction hypothesis in the second run to get $\pc^\dprime$ such that $\pc^\dprime \not\sqsubseteq \ladv$ and $\Gamma;\pc^\dprime \vdash c : \pc'$. We proceed by inspecting $\timestamp \in \textsf{dom}(\schedule)$.
            
                If $\timestamp \not\in \textsf{dom}(\schedule)$ then by definition of \textsf{send} we have $\beta_1 = \epsilon$ and we are done.
                
                If $\schedule(\timestamp)=\ell$ we consider each direction
                \begin{description}
                    \item[$(\rightarrow)$]
                        We have by Lemma \ref{lemma:sending_preserves_equivalence} have
                        $\textsf{send}(\outenv_1,\ell)=(\beta_1,\outenv'_1)$ and
                        $\textsf{send}(\outenv_2,\ell)=(\beta_2,\outenv'_2)$ such that $\outenv'_1 \ladvequiv \outenv'_2$ and $\projevent{\beta_1}{\ladv}=\projevent{\beta_2}{\ladv}$
                        and hence
                        \[
                            \xAngled{\xangled{{\normalfont\texttt{stop}},m_2,\inenv_2},\outenv_2,\schedule,\timestamp}
                            \rightarrowdbl_{(\timestamp:\epsilon,\beta_2)}
                            \xAngled{\xangled{{\normalfont\texttt{stop}},m_2,\inenv_2},\outenv'_2,\schedule,\timestamp'}
                        \]
                        and we are done.
                    \item[$(\leftarrow)$] 
                        As $\beta_2 \neq \epsilon$, the anchor-configuration steps \textsc{G-Stop} and we apply Lemma \ref{lemma:sending_preserves_equivalence} in both runs to get $\projevent{\beta_1}{\ladv} = \projevent{\beta_2}{\ladv}$ and $\outenv'_1 \ladvequiv \outenv'_2$ and we are done.
                \end{description}
             
                \item \case{\textsc{G-Block}}
                Immediate by Lemma \ref{lemma:sending_preserves_equivalence}.
                
                \item \case{\textsc{G-Stop}}
                Immediate by Lemma \ref{lemma:sending_preserves_equivalence}.
                
                \item \case{No step}
                It must be that $\timestamp \notin \textsf{dom}(\schedule)$ and hence by definition of \textsf{send} have $\beta_1 = \epsilon$ and we are done.
            \end{description}
        
        \item \case{\textsc{G-Stop}}
            We have $\alpha_1 = \epsilon$ and $m_1 = m'_1$ and $\inenv_1 = \inenv'_1$ and $\outenv_1 = \outenv'_1$ and $\schedule = \schedule'$. We consider the possible steps of the second run.
            
            \begin{description}
                \item \case{\textsc{G-Step}}
                We will show that condition 2 is satisfied. To this end, we first apply the induction hypothesis in the second run to get $\pc^\dprime$ such that $\pc^\dprime \not\sqsubseteq \ladv$ and $\Gamma;\pc^\dprime \vdash c : \pc'$. We proceed by inspecting $\timestamp \in \textsf{dom}(\schedule)$.
            
                If $\timestamp \not\in \textsf{dom}(\schedule)$ then by definition of \textsf{send} we have $\beta_1 = \epsilon$ and we are done.
                
                If $\schedule(\timestamp)=\ell$ we consider each direction
                \begin{description}
                    \item[$(\rightarrow)$]
                        We have by Lemma \ref{lemma:sending_preserves_equivalence} have
                        $\textsf{send}(\outenv_1,\ell)=(\beta_1,\outenv'_1)$ and
                        $\textsf{send}(\outenv_2,\ell)=(\beta_2,\outenv'_2)$ such that $\outenv'_1 \ladvequiv \outenv'_2$ and $\projevent{\beta_1}{\ladv}=\projevent{\beta_2}{\ladv}$
                        and hence
                        \[
                            \xAngled{\xangled{{\normalfont\texttt{stop}},m_2,\inenv_2},\outenv_2,\schedule,\timestamp}
                            \rightarrowdbl_{(\timestamp:\epsilon,\beta_2)}
                            \xAngled{\xangled{{\normalfont\texttt{stop}},m_2,\inenv_2},\outenv'_2,\schedule,\timestamp'}
                        \]
                        and we are done.
                    \item[$(\leftarrow)$] 
                        As $\beta_2 \neq \epsilon$, the anchor-configuration steps \textsc{G-Stop} and we apply Lemma \ref{lemma:sending_preserves_equivalence} in both runs to get $\projevent{\beta_1}{\ladv} = \projevent{\beta_2}{\ladv}$ and $\outenv'_1 \ladvequiv \outenv'_2$ and we are done.
                \end{description}
             
                \item \case{\textsc{G-Block}}
                Immediate by Lemma \ref{lemma:sending_preserves_equivalence}.
                
                \item \case{\textsc{G-Stop}}
                Immediate by Lemma \ref{lemma:sending_preserves_equivalence}.
                
                \item \case{No step}
                It must be that $\timestamp \notin \textsf{dom}(\schedule)$ and hence by definition of \textsf{send} have $\beta_1 = \epsilon$ and we are done.
            \end{description}
    \end{description}
\end{IEEEproof}

\subsection{Soundness of security type system} \label{appendix:soundness}

\externalinternalknowledge*
\begin{IEEEproof}
Given an $\lconfig_2$ s.t. $\lconfig \ladvequiv \lconfig_2$ and unfolding the definitions of attacker knowledge and internal knowledge we must show
\[
    (\filtertraceinternal{\tau}{\ladv}) = (\filtertraceinternal{\tau_2}{\ladv})
    \implies
    (\filtertrace{\tau}{\ladv}) = (\filtertrace{\tau_2}{\ladv})
\]
The proof is by induction on trace $\tau$.

\begin{description}
    \item \basecase{$\tau = \epsilon$}
    Immediate.
    
    \item \inductivecase{$\tau = \tau' \cdot (\timestamp: \alpha, \beta)$}
    We have $\tau_2 = \tau'_2 \cdot (\timestamp_2: \alpha_2, \beta_2)$
    and have IH
    \[
        (\filtertraceinternal{\tau'}{\ladv}) = (\filtertraceinternal{\tau'_2}{\ladv})
        \implies
        (\filtertrace{\tau'}{\ladv}) = (\filtertrace{\tau'_2}{\ladv})
    \]
    By assumption, we have $(\filtertraceinternal{\tau' \cdot (\timestamp: \alpha, \beta)}{\ladv}) = (\filtertraceinternal{\tau'_2 \cdot (\timestamp_2: \alpha_2, \beta_2)}{\ladv})$ and hence $\projeventinternal{\alpha}{\ladv} = \projeventinternal{\alpha_2}{\ladv}$ and $\projevent{\beta}{\ladv} = \projevent{\beta_2}{\ladv}$ and we consider the two results of the internal trace filtering:
    \begin{description}
        \item \case{$(\filtertraceinternal{\tau}{\ladv}) = (\filtertraceinternal{\tau'}{\ladv})$}
        By definition of internal trace filtering have $\projeventinternal{\alpha}{\ladv} = \projeventinternal{\beta}{\ladv} = \epsilon$, and hence definition of trace filtering we have $(\filtertrace{\tau}{\ladv}) = (\filtertrace{\tau'}{\ladv})$ and $(\filtertrace{\tau_2}{\ladv}) = (\filtertrace{\tau'_2}{\ladv})$ and we are done by IH.
        
        \item \case{$(\filtertraceinternal{\tau}{\ladv}) \neq (\filtertraceinternal{\tau'}{\ladv})$}
        We have $\timestamp = \timestamp_2$ and hence $(\timestamp: \projeventinternal{\alpha}{\ladv},\projevent{\beta}{\ladv}) = (\timestamp_2: \projeventinternal{\alpha_2}{\ladv},\projevent{\beta_2}{\ladv})$. From this, we trivially get $(\timestamp: \epsilon,\projevent{\beta}{\ladv}) = (\timestamp_2: \epsilon,\projevent{\beta_2}{\ladv})$ and are done by definition of trace filtering and IH.
    \end{description}
\end{description}
\end{IEEEproof}

\soundness*
\begin{IEEEproof}
By Definition \ref{definition:timing-sensitive_noninterference} and unfolding of program configuration $\lconfig$ we are given a program $c$, initial memory $m$ and initial input environment $\inenv$ such that
\[
    \xAngled{\xangled{c,m,\inenv},\outenv_\textit{init},\schedule_\textit{init},0}
    \rightarrowdbl^*_{\tau \cdot \gamma}
    \xAngled{\xangled{c',m',\inenv'},\outenv',\schedule',\timestamp'}
\]
and must show that for any $\ladv$ we have
\[
    k(c,m,\inenv,\tau \cdot \gamma,\ladv) \supseteq k(c,m,\inenv,\tau,\ladv)
\]

Let $(m_2,\inenv_2) \in k(c,m,\inenv,\tau,\ladv)$ and $\ladv$ be given. Unfolding the definition of attacker knowledge we have $\tau_2$ such that $(\filtertrace{\tau}{\ladv}) = (\filtertrace{\tau_2}{\ladv})$ and
\[
    \xAngled{\xangled{c,m_2,\inenv_2},\outenv_\textit{init},\schedule_\textit{init},0}
    \rightarrowdbl^*_{\tau_2}
    \xAngled{\xangled{c'_2,m'_2,\inenv'_2},\outenv'_2,\schedule'_2,\timestamp'_2}
\]
We must show that
\[
    \xAngled{\xangled{c,m_2,\inenv_2},\outenv_\textit{init},\schedule_\textit{init},0}
    \rightarrowdbl^*_{\tau_2 \cdot \gamma_2}
    \xAngled{\xangled{c'_2,m'_2,\inenv'_2},\outenv'_2,\schedule'_2,\timestamp'_2}
\]
such that $(\filtertrace{\tau \cdot \gamma}{\ladv}) = (\filtertrace{\tau_2 \cdot \gamma_2}{\ladv})$.

We do this by showing a stronger property using internal trace filtering and applying Lemma \ref{lemma:external_internal_knowledge}. We show that if
\begin{align*}
    \xAngled{\xangled{c,m_1,\inenv_1},\outenv_\textit{init},\schedule_\textit{init},0}
    &\rightarrowdbl^n_{\tau}
    \xAngled{\xangled{c'_1,m'_1,\inenv'_1},\outenv'_1,\schedule',n}\\
    &\rightarrowdbl_{\gamma}
    \xAngled{\xangled{c^\dprime_1,m^\dprime_1,\inenv^\dprime_1},\outenv^\dprime_1,\schedule^\dprime,n+1}
\end{align*}
then one of the following holds
\begin{enumerate}
    \item[(1)] either
        \begin{align*}
            \xAngled{\xangled{c,m_2,\inenv_2},\outenv_\textit{init},\schedule_\textit{init},0}
            &\rightarrowdbl^n_{\tau_2}
            \xAngled{\xangled{c'_2,m'_2,\inenv'_2},\outenv'_2,\schedule',n}\\
            &\rightarrowdbl_{\gamma_2}
            \xAngled{\xangled{c^\dprime_2,m^\dprime_2,\inenv^\dprime_2},\outenv^\dprime_2,\schedule^\dprime,n+1}
        \end{align*}
        s.t. $(\filtertraceinternal{\tau \cdot \gamma}{\ladv})=(\filtertraceinternal{\tau_2 \cdot \gamma_2}{\ladv})$ and $m^\dprime_1 \ladvequiv m^\dprime_2$ and $\inenv^\dprime_1 \ladvequivprog \inenv^\dprime_2$ and $\outenv^\dprime_1 \ladvequiv \outenv^\dprime_2$
    \item[(2)] or there is $n' \leq n$
        \begin{align*}
            \xAngled{\xangled{c,m_2,\inenv_2},\outenv_\textit{init},\schedule_\textit{init},0}
            &\rightarrowdbl^{n'}_{\tau_2}
            \xAngled{\xangled{c'_2,m'_2,\inenv'_2},\outenv'_2,\schedule',n'}
        \end{align*}
        s.t. $(\filtertraceinternal{\tau \cdot \gamma}{\ladv})=(\filtertraceinternal{\tau}{\ladv})=(\filtertraceinternal{\tau_2}{\ladv})$ and
        $m^\dprime_1 \ladvequiv m'_2$ and $\inenv^\dprime_1 \ladvequivprog \inenv'_2$ and $\outenv^\dprime_1 \ladvequiv \outenv'_2$ and $\not\exists \timestamp' \in \textsf{dom}(\schedule'): \timestamp' \geq n$
        and there is $\pc_1,\pc_2$ such that $\pc_1 \not\sqsubseteq \ladv$ and $\pc_2 \not\sqsubseteq \ladv$ and $c'_1$ is well-formed w.r.t $\Gamma,\pc_1,\pc'$ and and $c'_2$ is well-formed w.r.t $\Gamma,\pc_2,\pc'$
\end{enumerate}
In each case, the equality of traces using internal trace filtering and Lemma \ref{lemma:external_internal_knowledge} gives us equality of traces using trace filtering as desired.

The proof is by induction on $n$

\begin{description}
    \item \basecase{$n=0$}
        Immediate by Definition \ref{definition:attacker_knowledge}, Definition \ref{definition:internal_knowledge}, Lemma \ref{lemma:inenvequiv_implies_inenvequivprog}, and Lemma \ref{lemma:noninterference_global}.
        
    \item \inductivecase{$n>0$}
        We apply the induction hypothesis and get
        \begin{enumerate}
            \item[(1)] either
                \begin{align*}
                    \xAngled{\xangled{c,m_2,\inenv_2},\outenv_\textit{init},\schedule_\textit{init},0}
                    &\rightarrowdbl^{n-1}_{\tau_2}
                    \xAngled{\xangled{c'_2,m'_2,\inenv'_2},\outenv'_2,\schedule',{n-1}}
                \end{align*}
                s.t. $(\filtertrace{\tau}{\ladv})=(\filtertrace{\tau_2}{\ladv})$ and $m'_1 \ladvequiv m'_2$ and $\inenv'_1 \ladvequivprog \inenv'_2$ and $\outenv'_1 \ladvequiv \outenv'_2$
            \item[(2)] or there is $n' \leq n-1$
                \begin{align*}
                    \xAngled{\xangled{c,m_2,\inenv_2},\outenv_\textit{init},\schedule_\textit{init},0}
                    &\rightarrowdbl^{n'}_{\tau_2}
                    \xAngled{\xangled{c'_2,m'_2,\inenv'_2},\outenv'_2,\schedule',n'}
                \end{align*}
                s.t. $(\filtertrace{\tau}{\ladv})=(\filtertrace{\tau_2}{\ladv})$ and
                $m'_1 \ladvequiv m'_2$ and $\inenv'_1 \ladvequivprog \inenv'_2$ and $\outenv'_1 \ladvequiv \outenv'_2$ and $\not\exists \timestamp' \in \textsf{dom}(\schedule'): \timestamp' \geq n'$
                and there is $\pc_1,\pc_2$ such that $\pc_1 \not\sqsubseteq \ladv$ and $\pc_2 \not\sqsubseteq \ladv$ and $c'_1$ is well-formed w.r.t $\Gamma,\pc_1,\pc'$ and and $c'_2$ is well-formed w.r.t $\Gamma,\pc_2,\pc'$
        \end{enumerate}
        
        If (1), then we are done by Lemma \ref{lemma:noninterference_global}.
        
        If (2), then by lemma \ref{lemma:high_steps_high_effects} and transitivity we have $m^\dprime_1 \ladvequiv m'_2$, $\inenv^\dprime_1 \ladvequivprog \inenv'_2$, and for $\gamma_1 = (\timestamp: \alpha_1,\beta_1)$ that $\projeventinternal{\alpha_1}{\ladv}=\epsilon$. Hence by Lemma \ref{lemma:high_update_preserves_equivalence} and transitivity we have $\outenv^\dprime_1 \ladvequiv \outenv'_2$ and $\schedule^\dprime = \schedule'$. We are done by applying Lemma \ref{lemma:preservation_well-formedness} in both runs.
\end{description}
\end{IEEEproof}

\fi

\end{document}